\documentclass{iopjournal}

\usepackage{amssymb}
\usepackage{amsmath}
\usepackage{amsthm}
\usepackage{empheq}
\usepackage{mdframed}
\usepackage{booktabs}
\usepackage{lipsum}
\usepackage{color}
\usepackage{psfrag}
\usepackage{pgfplots}
\usepackage{bm}
\usepackage{hyperref}
\usepackage{tikz}
\usepackage{esvect} 
\usepackage{bbold}
\usepackage[toc,page]{appendix}
\usepackage{float}
\usepackage{placeins} 

\usepgfplotslibrary{colorbrewer}
\pgfplotsset{width=8cm,compat=1.9}
\usetikzlibrary{arrows.meta}
\usetikzlibrary{angles}
\usetikzlibrary{calc}
\usepackage[mode=buildnew]{standalone}

\newcommand{\comment}[1]{}

\newcommand{\e}{e^-}

\newcommand{\RR}{\mathbb{R}}
\newcommand{\NN}{\mathbb{N}}

\newcommand{\EE}{\mathcal{E}}
\newcommand{\Scal}{\mathcal{S}}

\newcommand{\eV}{\text{eV}}

\newcommand{\cx}{\text{cx}}
\newcommand{\iz}{\text{iz}}
\newcommand{\eff}{\text{eff}}
\newcommand{\rad}{\text{rad}}

\newcommand{\av}{\mathbf v}
\newcommand{\avin}{\av_\text{in}}
\newcommand{\avout}{\av_\text{out}}

\renewcommand{\sc}{\text{sc}}
\newcommand{\ej}{\text{ej}}
\newcommand{\sd}{\text{sd}}
\newcommand{\vsc}{v_\sc} 
\newcommand{\avsc}{\av_\sc} 

\newcommand{\avej}{\av_\ej}

\newcommand{\Ein}{E_0}
\newcommand{\Esc}{E_\sc}
\newcommand{\Eej}{E_\ej}
\newcommand{\Osc}{\Omega_\sc}
\newcommand{\Oej}{\Omega_\ej}

\newcommand{\scx}{\sigma_\cx}
\newcommand{\si}{\sigma_I}
\newcommand{\srr}{\sigma_\text{RR}}

\newcommand{\sv}{\left\langle\sigma v\right\rangle}
\newcommand{\svi}{\sv_I}
\newcommand{\svr}{\sv_\text{R}}
\newcommand{\svcx}{\sv_\cx}
\newcommand{\Esv}{\left\langle E\sigma v\right\rangle}

\newcommand{\dsde}{\frac{d\sigma}{dE}}

\newcommand{\dsdedodo}{\frac{d^3\sigma}{dE d\Osc d\Oej}}

\newcommand{\effsi}{\si^\text{eff}}
\newcommand{\effdsde}{\left(\dsde\right)^\text{eff}}

\newcommand{\Eevw}{\frac{1}{2}m_e|\av-\aw|^2}

\newcommand{\Eev}{\frac{1}{2}m_e v^2}

\newcommand{\siEvw}{\si\left(\Eevw\right)}
\newcommand{\siEv}{\si\left(\Eev\right)}

\newcommand{\srrEv}{\srr\left(\Eev\right)}

\newcommand{\Max}[1]{f_{\mathcal{M}_{#1}}}

\newcommand{\vte}{v_{T_e}}
\newcommand{\vtn}{v_{T_N}}

\newcommand{\termf}[2]{\left(\frac{\partial f_{#1}}{\partial t}\right)_{#2}}
\newcommand{\termE}{\termf e{I, \text{in}}}
\newcommand{\termEe}{\termf e{I, \sc}}

\newcommand{\termej}{\termf e{I,\ej}}
\newcommand{\inlinetermf}[2]{\left(\partial f_{#1}/\partial t\right)_{#2}}
\newcommand{\inlinetermE}{\inlinetermf e{I, \text{in}}}
\newcommand{\inlinetermEe}{\inlinetermf e{I, \sc}}
\newcommand{\inlinetermej}{\inlinetermf e{I,\ej}}

\newcommand{\aw}{\mathbf w}
\newcommand{\au}{\mathbf u}
\newcommand{\ax}{\mathbf x}
\newcommand{\of}{\bar f}
\newcommand{\deltavpll}{\bar{v}_\pll}
\newcommand{\pll}{\parallel}
\newcommand{\Gcx}{G_{\cx}}
\newcommand{\Gr}{G_R}
\newcommand{\Gi}{G_I}
\newcommand{\termof}[2]{\left(\frac{\partial \of_{#1}}{\partial t}\right)_{#2}}
\newcommand{\termofE}{\termof e{I, \text{in}}}
\newcommand{\termofEe}{\termof e{I, \sc}}
\newcommand{\termofej}{\termof e{I, \ej}}
\newcommand{\lsd}{\lambda_\sd}

\newcommand{\Eminr}{E_{\min}^\text{RR}}
\newcommand{\Emincx}{E_{\min}^\cx}

\newcommand{\Eminiz}{E_{\min}^\iz}
\newcommand{\Emaxiz}{E_{\max}^\iz}

\newcommand{\deltau}{\bar{u}}

\begin{document}

\articletype{Paper}

\title{Plasma--neutral collision operators for hydrogen isotopes:
	kinetic and gyrokinetic descriptions, comparison to fluid}

\author{Matthieu Protais$^{1,*}$\orcid{0009-0003-5852-7446},
	Guilhem Dif-Pradalier$^1$\orcid{0000-0003-4869-7049},
	Yanick Sarazin$^1$\orcid{0000-0003-2479-563X},
	Peter Donnel$^1$\orcid{0000-0002-6669-416X},
	Virginie Grandgirard$^1$\orcid{0000-0001-7821-9107},
	Yannick Marandet$^2$\orcid{0000-0002-9970-538X},
	Srikanth Sureshkumar$^1$\orcid{0009-0007-6426-9286}
	and Naren Varadarajan$^1$\orcid{0009-0001-5887-2587}
}

\affil{$^1$CEA, IRFM, 13108 Saint-Paul-lez-Durance, France}

\affil{$^2$Aix-Marseille Univ., CNRS, PIIM, 13397 Marseille, France}

\affil{$^*$Author to whom any correspondence should be addressed.}

\email{matthieu.protais@cea.fr}

\keywords{Plasma–neutral interactions, kinetic theory, gyrokinetic theory, inelastic collisions}

\begin{abstract}
	Plasma-neutral interactions modify ion and electron velocity distributions,
	potentially affecting plasma instabilities. In this work we derive ionization,
	recombination and charge exchange operators for hydrogen-like atoms in a fully
	kinetic framework, revisiting often-used assumptions of linearity and
	Maxwellianity, common in usual fluid-based approaches.
	We first present a comprehensive derivation of plasma-neutral interaction
	terms in the kinetic and gyrokinetic frameworks. The model naturally retains
	non-Maxwellian features of both the plasma and neutral distribution functions.
	By explicitly accounting for velocity-dependent differential cross-sections,
	the model captures the full kinetic impact of such inelastic collisions
	on the complete distribution functions of ions and electrons, and not only
	through their first moments. We include energy losses due to excitation,
	multi-step ionization and radiative recombination. We show that the fluid
	moments of the kinetic model reduce to standard fluid source terms under the
	appropriate assumptions of Maxwellianity and linearity, and specifically
	discuss deviations from fluid predictions over a range of tokamak-relevant conditions.
	Substantial discrepancies are observed, even for Maxwellian distributions.
	For charge exchange, the kinetic model yields momentum and energy exchange rates
	up to twice those predicted by fluid descriptions. We furthermore show that
	a fluid treatment of ionization greatly overestimates the mean kinetic energy
	of the produced electrons. These results pave the way for realistic kinetic
	interactions between a kinetic or gyrokinetic turbulent plasma and neutrals,
	particularly in the edge and scrape-off layer.
\end{abstract}

\section{Introduction}
\label{Sec:intro}

In recent years, due to the increase in computational power, kinetic and gyrokinetic
codes have started to turn their attention to the edge and scrape-off layer (SOL) of
magnetic fusion devices, where plasma transport simulation had previously been
accessible only through fluid approximations
\cite{becoulet_2026, hakim_2020, ku_2009, michels_2022}.
In this framework, the questions of kinetic effects due to the presence of neutral
atoms in the plasma become relevant, both through their possible interplay with
turbulence and as a realistic source of particles in the edge and SOL. These effects have
a direct impact on the plasma exhaust problem and in particular on plasma \emph{detachment},
\textit{i.e.} when the plasma is not in direct contact with the wall anymore,
and is separated from it by a layer of neutrals. This is one of the key regimes
envisioned for heat flux mitigation \cite{pitts_2019}.
This regime occurs when the plasma density close to the wall is high enough for the
recycling to be the dominant boundary condition, leading to a loss of plasma kinetic
(most importantly), and thermal (to a lesser extent) energy, ultimately reaching temperatures
allowing recombination. Similarly, controlled impurity seeding is often used
to increase radiation loss and reduce heat flux to the divertor plates. This is
particularly the case in the \emph{X-point radiator} regime, or XPR, which has been
heavily studied in the past years \cite{eder_2025, pan_2023, rivals_2024},
in which a stable radiation front is observed around the X-point, coming from the
accumulation of impurities. These important regimes of operation rely on the
interaction between the charged species of the plasma and neutral atoms.

Neutrals have an impact on the plasma in several non-trivial ways. Firstly,
they act as a source and/or a sink of density, momentum and energy (in the reminder
of the article, ``source'' will often be used, whether it actually refers to a source
or a sink). For instance, ionization is a net source of density for electrons and ions,
as well as a net sink of both momentum and energy for the electrons.
Secondly, the interaction between plasma species and neutrals
modifies the shape of the electron and ion distribution functions, driving them away
from Maxwellians. These modifications occur in regions of the velocity space
where it can affect the drive of
turbulent instabilities, as the latter often involve resonant interactions
between plasma species and electromagnetic waves. As far as we know, the issue of
the possible kinetic impact of plasma-neutral interactions on turbulence is an
unknown territory. Depending on plasma conditions, plasma-neutral interaction
time scales can be comparable to elastic collision time scales, leading
to non-trivial effects in velocity space.
This, however, is highly dependent on both plasma conditions and neutral density.
For ITER envisioned conditions in the edge (region of the core close to the separatrix),
say $T_e\sim 10^3\eV$ and $n_N\sim 10^{18}\text{m}^{-3}$ \cite{sureshkumar_2024},
the ionization collision frequency can be of the order of
$\nu_I=n_N\svi(T_e)\sim 10^4\text{s}^{-1}$. Compared to this, the electron-electron
elastic collision frequency $\nu_{ee}\sim n_e m_e^{-1/2} T_e^{-3/2}$
in these plasma conditions can be of similar orders of magnitude.
Moreover, if the shape in velocity space of these distribution functions affects
turbulence growth, it also feeds back non trivially on the atomic (inelastic) reactions
that occur between the plasma and the neutrals. For instance, ionization being
a threshold reaction (an electron with less energy than the ionization energy of the
neutral cannot ionize it), taking into account the actual -- and possibly non-Maxwellian --
energy distribution of electrons is important to get a accurate estimate of the actual
ionization rate.

In most models of plasma transport, the dynamics of the plasma and of the neutrals only
exchange information through their first three fluid moments.
A neutral solver is used to compute the transport of neutrals for some
given plasma conditions (density, momentum and temperature). The resulting
neutral fluid moments are then used to compute plasma source and sink terms,
incorporated in a plasma solver. The plasma solver then solves the plasma dynamics,
either in a fluid way \cite{bufferand_2019, giacomin_2022, stegmeir_2018, wiesen_2015}
or in a gyrokinetic way \cite{grandgirard_2016, ulbl_2025}, tracking its evolution
on relevant time scales. In the past several years, a number of codes have therefore
been developed to solve the transport of neutrals. High fidelity kinetic
solvers, using Monte Carlo methods to statistically simulate trajectories of atoms
and molecules, like EIRENE \cite{reiter_2005} or DEGAS2 \cite{stotler_1994},
remains computationally expensive. Some efforts have been done to develop reduced models
\cite{helander_1994, horsten_2017}, using fluid hypotheses for the neutral species.
Such codes are able to predict the localisation and temperature of multiple species
of neutrals, usually under restrictive hypotheses on the temperature regime,
such as assuming that the dynamics of neutrals is dominated by the effect of charge exchange
interactions with the ions. Even in cases in which the neutral dynamics is solved within
a kinetic description, neutral and plasma species interactions are mostly solved
within the fluid framework, \textit{i.e.} via their fluid moments only. In that respect,
the source and/or sink terms remain fluid in nature, even though the particle dynamics
are treated kinetically.

All of the previously cited neutral solvers assume the background plasma to
be Maxwellian to solve for the neutrals dynamics, and only incorporate the effect
of neutrals on the plasma evolution through the first three moments (density,
momentum and energy). The only exception, to the authors' knowledge, is the BIT1
code \cite{tskhakaya_2010}, able to handle kinetically the inelastic collisions
between particles \cite{tskhakaya_2023}, in a particle-in-cell (PIC) framework.
This paper aims at filling this gap by developing a set of kinetic
and gyrokinetic (grid based) operators, taking into account the full distribution
functions of all considered species without making any assumption about their shape,
in particular as to whether or not they are Maxwellian. These proposed operators produce
kinetic and gyrokinetic source terms in the Vlasov equation for three atomic reactions
usually dominant in tokamak plasmas: electron impact ionization, radiative recombination
and resonant charge-exchange. This work focuses on hydrogen-like atoms (hydrogen
isotopes), and highlights the main challenges in generalizing to other species.

The paper is structured as follows. Sections~\ref{Sec:kin_iz} and \ref{Sec:kin_recCX}
provide a derivation of kinetic operators for the three reactions of interest,
the former focusing on ionization, and the latter on recombination and charge exchange.
Section~\ref{Sec:gyro_derivation} expresses the newly derived terms in the gyrokinetic
framework. A comparison with fluid models is presented in Section~\ref{Sec:comp_fluid},
both in terms of fluid moments and kinetic effects, before concluding in
Section~\ref{Sec:conclusion}.

\section{Kinetic treatment of ionization}
\label{Sec:kin_iz}

In this section, we are interested in the ionization reaction, and its effect on
both ion and electron distribution functions. We first review the fundamental and
well-known aspects of this reaction. In particular, this pedagogical introduction
establishes the concepts and notation required for the subsequent kinetic derivation,
and allows for an exposition of our own simplifying hypotheses, used throughout the
derivation. As these concepts are at the confluence of different communities (plasma
physics and atomic physics), there is a strong pedagogical value in exposing the
atomic physics basis to enable their usage in plasma physics. In order not to confuse
the notation for ions and for ionization we use the subscript $i$ when referring to
the ions species, and $I$ when speaking about ionization. Moreover, we distinguish
two notations, using parenthesis to indicate the dependence of a quantity to another
one (e.g. $\si(E)$), and square brackets to group terms (e.g. $[1+\lambda]\av$).

\subsection{Preliminaries on the ionization of a neutral}
\label{Sec:iz_def}

Consider a population of hydrogen atoms. Even in the absence
of any plasma, depending on the temperature of the gas, each of the
atoms will be in a given excitation state. At steady-state, most of
them will be in the so-called \emph{ground state}. By convention, we set the
potential energy of a hydrogen atom in the ground state to be
$-E_I=-13.6\eV$. This population of atoms will exchange energy,
in the form of photons, through three main processes:
spontaneous emission, stimulated emission and photon absorption.
The probability of occurrence of each of these processes is given by
the so-called \emph{Einstein coefficients}. For instance,
$A_{j\to k}$ denotes the probability of spontaneous
emission to decay from a state $j$ to a state $k$. For more details we
refer e.g. to \cite{bransden_atoms_molecules}.
A hydrogen atom can be in an infinite but discrete number of
excited states, with energy levels $E_k=-E_I/k^2$,
$k=1,\dots,\infty$.
If the atom reaches an energy level greater than $E_\infty=0$
it is ionized and the electron is not bound anymore to the
nucleus. The energy of the electron lies then in a continuous spectrum.

\begin{figure}
	\centering
	\includegraphics[width=0.5\textwidth]{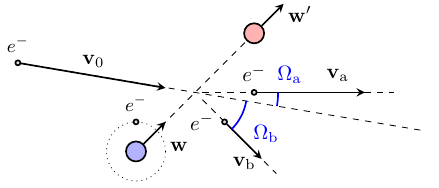}
	\caption{General case of an ionization event: an incident electron
		(velocity $\av_0$) collides with a hydrogen atom (velocity $\aw$) an ejects
		the bound electron. The result is a proton (velocity $\aw'$) and two
		outgoing electrons (velocity $\av_a$ and $\av_b$). We use the color
		blue for a proton when bound to an electron.
		The same convention is used for subsequent figures.}
	\label{Fig:general_iz}
\end{figure}

In the case of a tokamak plasma, the neutrals are surrounded
by a plasma background, whose electrons (mainly) can transfer
energy to the neutrals, exciting them to a higher energy state.
If the energy transferred from the electron to the neutral
is large enough, this can lead to the ionization of the atom,
ejecting the bound electron. This is called \emph{electron
	impact ionization},
\begin{equation} \label{ionization_excited_Eq}
	H(k) + e^- \to H^+ + e^- + e^-,
\end{equation}
with $H(k)$ denoting a neutral in the $k$th excited state.
Figure~\ref{Fig:general_iz} shows a simplified view of what happens during
such a process, hydrogen atoms being drawn using Bohr's model.
For explanation purpose we use a classical depiction of both the electrons
and the atoms, even though the physics at play, such as cross sections,
are computed using a quantum theory.
An incident electron with a velocity $\av_0$ approaches a hydrogen atom,
that has a velocity $\aw$. The electric fields associated to the incident electron
and to the trapped electron interact, and some energy is transferred from
the incident electron to the trapped one. In this case,
the energy of the trapped electron is greater than the ionization
energy of the atom, and the trapped electron escapes the
nucleus potential well.
The result of the collision is the slight alteration
of $\aw$, as well as the presence of now two electrons with velocities $\av_\text{a}$
and $\av_\text{b}$. These two electrons are emitted at a given solid angle,
$\Omega_a$ and $\Omega_b$ respectively.
By convention, these two angles are relative to the incident velocity vector.

In this framework, the outcome of an ionization event is
fully determined by the so-called \emph{triply differential cross
	section}: in energy and in scattering angles for both of the
outgoing electrons. In general, this differential cross section
$d^3\sigma/dEd\Omega_a d\Omega_b$ depends
on the incident energy, as well as on the angles of ejection,
the energy transfer from one electron to the other, and
the excited state of the neutral. When nothing is explicitly
indicated, the cross sections are taken to be the ground state ones.
For more details about the definition of differential cross sections we refer to
\cite{ehrhardt_differential_1986}.

\begin{figure}
	\centering
	\includegraphics[width=0.5\textwidth]{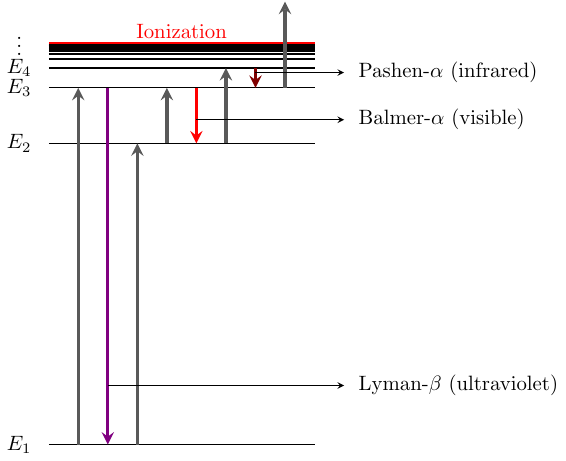}
	\caption{Arbitrary example of the multi-step ionization of a neutral}
	\label{Fig:multi_step_ionization}
\end{figure}

During the lifespan of a neutral, before it is ionized,
it can experience multiple changes in its energy level.
As an example, Figure~\ref{Fig:multi_step_ionization}
shows the possible evolution of the energy level of a given atom.
In the considered case, the hydrogen atom is first
excited by the collision with an electron,
but radiates out all the energy soon after, emitting a photon
in the Lyman series, that brings it back to the ground state.
It is then excited once more, and has no time
to de-excite before the next collision with an electron.
After some back-and-forth it is finally ionized.
It is to be noted that this multi-step process occurs at fast time scales.
Typically, the lifetime of excited states of a neutral is of the order
of $10^{-8}$s \cite{summers_2006}. This dynamics has two major effects.
Firstly, every time a neutral de-excites, it releases a photon that can be lost
as radiation, leading to the cooldown of the overall system.
Secondly, depending on the average energy level of the neutrals,
the energy needed to ionize them changes, and with it the
ionization cross section. The two effects will be discussed in
Section~\ref{Sec:iz_CRM}.

\subsection{Assumptions and simplifications}
\label{Sec:iz_ass}

We now present the assumptions chosen in this work to simplify the handling
of ionization. As shown in Figure~\ref{Fig:general_iz}, after
an ionizing collision, one can measure two outgoing electrons.
The three quantities of interest are $\Omega_a$, $\Omega_b$ and
the energy of one of the outgoing electrons (which determines the energy of the other).
When deriving the distribution in energy of the outgoing electrons
\cite{brauner_1989}, or experimentally measuring it
\cite{ehrhardt_differential_1986}, one
important feature is the separation of the electrons in two classes, one
with high energy and one with low energy.
This is why, conventionally, one separates the outgoing electrons
in two groups, sometimes called ``fast'' and ``slow'', ``hot'' and ``cold'',
or ``primary'' and ``secondary''. We choose to stick to the terminology
presented in \cite{ehrhardt_differential_1986}, calling the most energetic
the ``scattered'' one, and the other one the ``ejected'' one.
The velocities and solid angles shown in Figure~\ref{Fig:general_iz}
thus become $\avsc$, $\Osc$ and $\avej$, $\Oej$ accordingly.
It is to be noted that ionization being a quantum process,
the two produced electrons are indistinguishable,
\textit{i.e.} no experiment could tell which one was the incident
electron and which one was bound to the nucleus. Therefore,
the classification in two groups is only a terminology and
does not reflect a physics splitting.
In further simplifications, see Eq.~\eqref{eq:ass_on_dsdedodo},
these two groups are assumed to have different angle distributions,
creating an artificial separation between them.
This is the main limitation of this model, but
simplifies greatly the problem at hand.
Future work could alleviate this restriction by considering
the full triply differential cross section and its dependence
in the outgoing energy of the electrons.
Generalization is fairly straightforward, provided the triply
differential atomic data exists, but will not be considered in the
remaining of this article.

\begin{figure}
	\centering
	\includegraphics[width=0.7\textwidth]{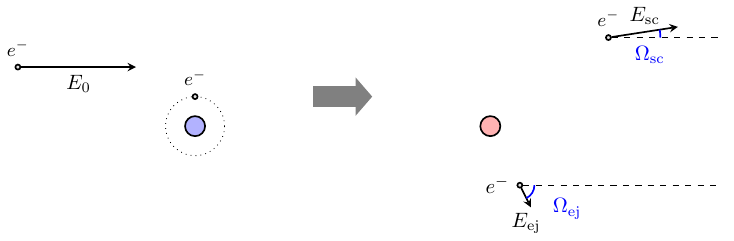}
	\caption{Electron impact ionization of an hydrogen atom in the center of mass frame,
		assumed to be the neutral frame of reference}
	\label{Fig:iz_com}
\end{figure}

We furthermore place ourselves in the reference frame of the neutral.
In particular, all kinetic energies will be calculated in the reference
frame of the center of mass, which, at leading order, coincides
with the neutral's one given that $m_N\gg m_e$.
We simplify the notations by setting
$E_0=\frac{1}{2}m_e|\av_0-\aw|^2$ for the incident energy,
$\Esc=\frac{1}{2}m_e|\avsc-\aw|^2$ for the energy of the scattered
electron and $\Eej=\frac{1}{2}m_e|\avej-\aw|^2$ for the ejected one.
We note here that by conservation of energy these quantities
are linked by the relation
\begin{equation}
	\Ein=E_I^*+\Eej+\Esc,
	\label{eq:conservation_energy_iz}
\end{equation}
with $E_I^*$ the energy needed to ionize the neutral,
\textit{i.e.} $E_I^*=-E_k$ for an atom in the $k$th excited state (remember
that by convention $E_k<0$ and $E_1=-E_I$).
The situation is represented in Figure~\ref{Fig:iz_com}.

In this context, the differential cross section defined in
Section~\ref{Sec:iz_def} becomes
\begin{equation}
	\dsdedodo(\Ein, \Esc, \Osc, \Oej),
	\label{eq:dsdedodo}
\end{equation}
and represents the probability that the scattered electron ends
up with energy $\Esc$, while being scattered by an angle $\Osc$
and that the ejected electron is emitted at a solid angle $\Oej$.
It does not seem to exist any convention on the energy taken
for the differential cross section, \textit{i.e.} it can be equivalently
$\Esc$ \cite{czyzewski_1990, ehrhardt_differential_1986},
$\Eej$ \cite{oda_energy_1975, purohit_electron_2024, shyn_doubly_1992},
or even $\Eej+E_I^*$ \cite{rax_physique_2005}.
As they are all equivalent we choose the first one.

When integrating Eq.~\eqref{eq:dsdedodo} in both
$\Osc$ and $\Oej$, one obtains the singly differential cross
section in energy, $d\sigma/dE$, which depends on $(\Ein,\Esc)$ only.
It represents the probability for an incident electron
with energy $\Ein$ to undergo an ionizing collision, be scattered
with final energy $\Esc$ and lead to an ejected electron
of energy $\Eej=E_0-E_I^*-\Esc$.
This cross section equals zero outside the
allowed range of energy. In practice it means that
\begin{equation}
	\dsde(\Ein,\Esc)\neq 0 \iff 0 \le \Esc \le \Ein - E_I^*.
	\label{eq:allowed_range_energy_iz}
\end{equation}
For instance, when the energy range of an integral is not specified,
it is assumed to be the support of $d\sigma/dE$.
We emphasize the fact that Eq.~\eqref{eq:allowed_range_energy_iz}
is not independent of $\Eej$. Indeed, to a given pair $(\Ein, \Esc)$
corresponds only one energy $\Eej$ given by $\Eej = \Ein - \Esc - E_I^*$,
see Eq.~\eqref{eq:conservation_energy_iz}.
For more details about the expression of these cross sections, we refer the
reader to Appendix~\ref{Annex:cross_sections}.

Very importantly we have the following relation between the
various introduced cross sections:
\begin{equation}
	\int d\Osc \int d\Oej \int_0^{\Ein} d\Esc\; \dsdedodo(\Ein,\Esc,\Osc,\Oej)=
	\int_0^{\Ein} d\Esc\; \dsde(\Ein,\Esc)=\si(\Ein).
	\label{eq:int_dsde_eq_sigma}
\end{equation}
Note that we chose to use $[0,\Ein]$, and not $[0,\Ein-E_I^*]$,
as the integration range, for the sake of simplicity.
The two are equivalent in virtue of Eq.~\eqref{eq:allowed_range_energy_iz}.

\begin{figure}
	\centering
	\includegraphics[width=0.8\textwidth]{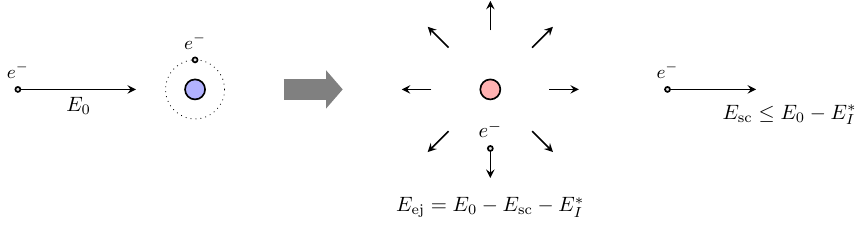}
	\caption{Final hypotheses made in this article for the
		ionization of an hydrogen atom. The scattered electron
		is not deflected (its velocity has the same direction
		than the incident electron), and the ejected one is emitted isotropically.}
	\label{Fig:ionization}
\end{figure}

We now introduce the main simplification of this work.
Theoretical studies \cite{shyn_doubly_1992}, and experimental
measurements \cite{ehrhardt_differential_1986},
both show a strong dependence of the angular distribution
of electrons to the energy of the outgoing electron.
More precisely, the more energetic the outgoing electron, the closer
to the forward direction it is emitted.
Conversely, the slow ejected electrons from an ionizing
collision are emitted more isotropically. This is clear in
\cite{mott_theory_1987}, Fig.~40.
It has been proposed \cite{janev_1995}, and been
used \cite{tskhakaya_2023}, to model the angle distribution
of outgoing electrons with a parameter increasing the anisotropy
at high energies.
In this work, we consider $\Osc$ to be zero, and the triply differential
cross section independent of $\Oej$.
This assumption becomes increasingly good at high impact energy regimes,
\textit{i.e.} for fast incident electrons.
In that case the triply differential cross section becomes
\begin{equation}
	\dsdedodo(\Ein, \Esc, \Osc, \Oej)=\frac{1}{4\pi}\delta_\mathbf{0}(\Osc)\dsde(\Ein,\Esc).
	\label{eq:ass_on_dsdedodo}
\end{equation}
We used $\delta_\mathbf{0}(\Osc)$ to signify that the two vectors
are collinear, $\av_0\parallel\avsc$ (\textit{i.e.} we neglect the scattering).
This assumption simplifies the impact of ionization on
the distribution function of electrons.
Figure~\ref{Fig:ionization} shows the final hypotheses made is this paper.
The presence of these three electrons during the reaction
naturally drives us to decompose the kinetic term describing the reaction
in three parts, namely
\begin{equation}
	\termf eI = - \termE + \termEe + \termej,
	\label{eq:sep_term_dfe}
\end{equation}
with ``$\text{in}$'' referring to the incident electron,
``$\sc$'' the scattered one, and ``$\ej$'' the ejected one.
As the incident electron and the scattered share the same
velocity direction, only with $v_0>v_\sc$, we refer to the
sum $-\inlinetermE+\inlinetermEe$ as the \emph{slowdown} of
the incident electron, which is again only terminology and does
not reflect the actual physical process due to the indistinguishable nature
of the two resulting electrons.

In the next sections we derive the kinetic terms taking
ionization into account in the Vlasov equation.
In Section~\ref{Sec:iz_ground_state},
we start by assuming the neutral to be in the ground state.
This hypothesis will then be alleviated in Section~\ref{Sec:iz_CRM}.

\subsection{Simplified framework: ionization from the ground state}
\label{Sec:iz_ground_state}

In this section, we derive the kinetic operator for an electron impact
ionization, under the hypothesis that the reacting neutral is in the ground
state. In our previous notations, it means that $E_I^*=E_I$, and
Eq.~\eqref{ionization_excited_Eq} becomes
\begin{equation} \label{ionization_Eq}
	H + e^- \to H^+ + e^- + e^-,
\end{equation}
with $H$ denoting a neutral in ground state, $H(1)$.
This assumption will be removed in Section~\ref{Sec:iz_CRM}.

\subsubsection{Impact on the ion distribution function}
\label{Sec:ionization_on_ions}

Consider first the effect of an ionization event like in Figure~\ref{Fig:general_iz}
on the ion species. Due to the large proton to electron mass ratio,
the proton momentum can be assumed to remain unchanged at leading order.
Therefore the velocity of the resulting ion is taken to be
the initial velocity of the neutral, $\aw'=\aw$, where the prime
denotes the state after the collision.
The derivation of $(\partial f_i/\partial t)_\text{I}$ is thus fairly straightforward.
An ionization collision will modify the value of the distribution function
at velocity $\aw$ if a hydrogen atom with velocity $\aw$ gets ionized.
Considering an atom at velocity $\aw$ in a cloud of electrons
with distribution function $f^t_e$,
the probability that it gets ionized during a time duration $\Delta t$ is
\begin{equation}
	p_N(\aw)=\Delta t\int d^3\av\; |\aw-\av|
	\siEvw f^t_e(\av).
	\label{eq:pn_iz}
\end{equation}
If at time $t$ the distribution function of ions is $f^t_i$, at time $t+\Delta t$ it will be
\begin{equation*}
	f^{t+\Delta t}_i(\aw) = f^{t}_i(\aw) + p_N(\aw) f^t_N(\aw).
\end{equation*}
Dividing by $\Delta t$ and taking the limit toward zero yields
\begin{equation}
	\termf {i}{I} (\aw)
	= f_N(\aw)\int d^3\av\; |\aw-\av|\siEvw f_e(\av).
	\label{eq:termf_iz_ion}
\end{equation}
These kinds of integral operators
are called \emph{Boltzmann operators} and will be found
everywhere throughout this work.
We now turn our attention to the electron species.

\subsubsection{Momentum loss of the incident electron}
\label{Sec:ionization_slowdown}

In this section we compute the slowdown term appearing
in Eq.~\eqref{eq:sep_term_dfe}, namely $-\inlinetermE+\inlinetermEe$.
Given an incident electron with energy $\Ein$,
for any ionization event, we should remove it from the distribution
function at that energy and add it back with energy $\Esc<\Ein-E_I$.
To compute $\inlinetermE$, similarly to the reasoning in
Eq.~\eqref{eq:pn_iz} with the ion species, we use the probability
for an electron with velocity $\av$ to undergo an ionization collision.
It yields
\begin{equation}
	\termE(\av) = f_e(\av)\int d^3\aw\; |\av-\aw|
	\si\left(\frac{1}{2}m_e|\av-\aw|^2\right)
	f_N(\aw)
	\label{eq:term_sd_sink}
\end{equation}
We note here that we use the full ionization cross section $\si$ defined
in Eq.~\eqref{eq:int_dsde_eq_sigma} and not the
differential one. Indeed, if an electron undergoes an
ionization event (no matter the energy lost) it should be removed from
the distribution function.

The term $\inlinetermEe$ should then add this colliding electron back
in the electron distribution function at another velocity whose probability
is given by the differential cross section in energy and angles.
To account for the absence of creation or annihilation of the electron,
the following condition should therefore be strictly satisfied:
\begin{equation}
	\int d^3\av_0\; \termE(\av_0)=\int d^3\avsc\; \termEe(\avsc).
	\label{eq:sd_no_density}
\end{equation}
Eq.~\eqref{eq:term_sd_sink} does not depend on $d\sigma/dE$, but we can use
Eq.~\eqref{eq:int_dsde_eq_sigma} to make his dependence explicit,
and the assumptions presented in Section~\ref{Sec:iz_ass},
namely Eq.~\eqref{eq:ass_on_dsdedodo}, to simplify it.
This new expression will reveal useful so as to enforce the relation
Eq.~\eqref{eq:sd_no_density}.
\begin{equation}
	\termE(\av_0) =\int d^3\aw\;|\av_0-\aw|
	\left[ \int d\Osc\int_0^{\Ein}d\Esc\;
		\delta_\mathbf{0}(\Osc)\dsde(\Ein,\Esc) \right]
	f_N(\aw)f_e(\av_0).
	\label{eq:1_expansion_termE}
\end{equation}
We recall here that
$\Ein=\frac{1}{2}m_e|\av_0-\aw|^2$ and $\Esc=\frac{1}{2}m_e|\avsc-\aw|^2$.
Note that in Eq.~\eqref{eq:1_expansion_termE}, the integral in
$\Oej$ present in Eq.~\eqref{eq:int_dsde_eq_sigma} canceled out with
the $1/4\pi$ factor in Eq.~\eqref{eq:ass_on_dsdedodo}.

Starting from Eq.~\eqref{eq:sd_no_density}, and injecting
Eq.~\eqref{eq:1_expansion_termE}, it is now a matter of
exchanging the order of integration
in order to switch from the incident electron lens
to the scattered electron one.
For the sake of compactness, we define a new dimensionless
quantity $\lsd$, as the ratio of velocities before
and after the collision, which reads
\begin{equation}
	\lsd=\dfrac{|\av_0-\aw|}{|\avsc-\aw|}.
	\label{eq:def_lambda_sd}
\end{equation}
Note that $\lsd^2$ is the ratio of energies $\Ein/\Esc$.
We refer to Appendix~\ref{Annex:kin_details} for the computations, which yield
\begin{equation}
	\termEe(\avsc)=
	m_e\int d^3\aw \;f_N(\aw)
	\int_1^\infty d\lsd\;
	\lsd^3|\avsc-\aw|^3
	\dsde(\lsd^2\Esc,\Esc)f_e\Big(\aw+\lsd[\avsc-\aw]\Big).
	\label{eq:term_sd_source}
\end{equation}

\begin{figure}
	\centering
	\includegraphics[width=0.6\textwidth]{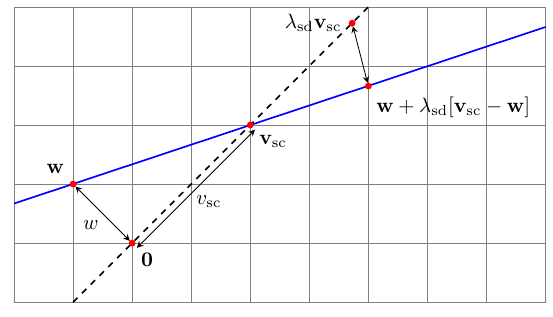}
	\caption{Approximation~\eqref{eq:term_sd_source_approx}
		for a two-dimensional velocity space, the blue line indicates
		the correct range of integration, the black dashed one the
		approximated one}
	\label{Fig:assumption_velocity}
\end{figure}

We introduce a last simplification, leveraging the fact
that the width of $f_e$ (typically the thermal velocity
of electrons in the case of a Maxwellian) is largely greater than the
thermal velocity of neutrals.
As neutrals are expected to have a lower temperature than electrons,
$\aw$ is typically orders of magnitude smaller than the thermal velocity
of electrons, since
\begin{equation*}
	\vtn=\sqrt{\frac{m_eT_N}{m_iT_e}}\vte.
\end{equation*}
Therefore, in the case where $\avsc$ is far from $0$
(typically $v_\sc>\vte$),
$\aw$ is negligible compared to $\avsc$,
and $\lsd$ is bounded (see Eq.~\eqref{eq:def_lambda_sd}).
We assume
\begin{equation*}
	f_e\Big(\aw+\lsd[\avsc-\aw]\Big)
	= f_e\left(\lsd \avsc + [1 - \lsd]\aw\right)
	\simeq f_e\left(\lsd \avsc\right).
\end{equation*}
This approximation breaks down at small $\avsc$,
typically when $\avsc-\aw$ approaches zero, as $\lsd$ grows like
$1/|\avsc-\aw|$.
This region of the velocity space is however not a problem
because $d\sigma/dE(\lsd^2\Esc, \Esc)\to 0$ when $\lsd\to 0$.
This allows us to write the following approximation for
Eq.~\eqref{eq:term_sd_source}:
\begin{equation}
	\termEe(\avsc)\simeq
	m_e\int d^3\aw \;f_N(\aw)
	\int_1^\infty d\lsd\;
	\lsd^3|\avsc-\aw|^3
	\dsde(\lsd^2\Esc,\Esc)f_e(\lsd\avsc).
	\label{eq:term_sd_source_approx}
\end{equation}
Figure~\ref{Fig:assumption_velocity} shows the approximation in a two-dimensional
velocity space. Note that the figure is deliberately exaggerated, in reality
$w$ is usually negligible compared with $\vsc$.
It is however to be remembered that this approximation introduces an
error in Eq.~\eqref{eq:sd_no_density} which should
be numerically corrected \textit{a posteriori}. Indeed, Eq.~\eqref{eq:sd_no_density}
should be enforced with the greatest precision.

Note that Eq.~\eqref{eq:term_sd_source} and \eqref{eq:term_sd_source_approx}
are well-defined and regular (in particular at $\avsc=0$) thanks to the fact
that $\lim_{\Esc \to 0} d\sigma/dE(\lsd^2 E_\sc, E_\sc) = 0,\ \forall \lsd>0$.

\subsubsection{Ejection of cold electrons}

We finally turn our attention to $\inlinetermej$,
which accounts for the cold electrons produced by ionization.
To modify the electron distribution function at a given velocity
$\avej$, cold electrons should be produced at velocity
$\avej$, corresponding to the energy $\Eej=\frac{1}{2}m_e|\avej-\aw|^2$.
This can only happen if an incident electron with energy
$\Ein=\frac{1}{2}m_e|\av_0-\aw|^2$ transfers an energy of $\Eej+E_I$ to the atomic
electron, and leaves with scattered energy $\Esc=\Ein-\Eej-E_I$.
We recall here that, in this section~\ref{Sec:iz_ground_state},
the neutrals are assumed to be in the ground state,
so that $E_I^* = E_I$. The generalization to excited states is addressed
in Section~\ref{Sec:iz_CRM}.
In addition, since one incident electron produces one ejected electron, we should have
\begin{equation}
	\int \termE(\av_0)\;d^3\av_0=\int \termej(\avej)\;d^3\avej.
	\label{eq:ej_no_density}
\end{equation}
As Eq.~\eqref{eq:sd_no_density}, Eq.~\eqref{eq:ej_no_density} should be enforced
with the greatest precision.

To derive Eq.~\eqref{eq:term_sd_source} we started
from Eq.~\eqref{eq:sd_no_density} and \eqref{eq:1_expansion_termE}
and made our way through exchanging the inner and outer integrals.
This allowed us to switch from a $\av_0$ point of view to a $\avsc$ one.
We now use the same technique to bring out an integral over $\avej$,
using the same steps as in Section~\ref{Sec:ionization_slowdown},
this time keeping the integral over $\Oej$ and not the one over $\Osc$,
setting $\Eej=\Ein-\Esc-E_I$.
The counterpart of Eq.~\eqref{eq:1_expansion_termE} for the ejected
electron reads
\begin{equation}
	\begin{aligned}
		\termE(\av_0)
		 & =\int d^3\aw\;|\av_0-\aw|
		\left[\frac{1}{4\pi}\int d\Oej\int_0^{\Ein}d\Esc\;
			\dsde(\Ein,\Esc) \right]
		f_N(\aw)f_e(\av_0)
		\\
		 & =\int d^3\aw\;|\av_0-\aw|
		\left[\frac{1}{4\pi}\int d\Oej\int_0^{\Ein}d\Eej\;
			\dsde(\Ein,\Ein-\Eej-E_I) \right]
		f_N(\aw)f_e(\av_0),
	\end{aligned}
	\label{eq:1_expansion_termej}
\end{equation}
where in the last equality we were allowed to replace
$d\Esc$ by $d\Eej$ without changing the integral bounds
by using the fact that
\begin{equation*}
	\Eej\notin \Big[0,\Ein-E_I\Big]
	\implies
	\dsde(\Ein,\Ein-\Eej-E_I)=0.
\end{equation*}
The derivation of $\inlinetermej$ is done in
Appendix~\ref{Annex:kin_details} and yields
\begin{equation}
	\termej(\avej)=
	\frac{m_e}{4\pi}
	\int d^3\aw\;f_N(\aw)
	\int_{|\av_0-\aw|\ge|\avej-\aw|}d^3\av_0\;
	\frac{|\av_0-\aw|}{|\avej-\aw|}
	\dsde(\Ein,\Ein-\Eej-E_I) f_e(\av_0).
	\label{eq:termej}
\end{equation}
We note that Eq.~\eqref{eq:termej} includes
an integral on the full velocity space in $\av_0$, in contrast to the support
of integration in Eq.~\eqref{eq:term_sd_source_approx} that is only on a line,
owning to the hypothesis in Eq.~\eqref{eq:ass_on_dsdedodo}.
Similarly to Eq.~\eqref{eq:term_sd_source}
and \eqref{eq:term_sd_source_approx}, Eq.~\eqref{eq:termej}
is well-defined and continuous. This time the proof is more
subtle and relies on the regularity of $f_N$ (namely
$f_N\in C^\infty_0(\RR^3)$). This result breaks down at lower
regularity of $f_N$\footnote{For instance, Eq.~\eqref{eq:termej}
	is not bounded at $\av=\mathbf{0}$ for $f_N=\delta_\mathbf{0}$.}.

\subsection{Excitation and multi-step ionization: energy lost by radiation}
\label{Sec:iz_CRM}

In this section, we alleviate the assumption of ground state for the
neutral, therefore generalizing the reasoning followed in
Section~\ref{Sec:iz_ground_state}.
As introduced in Section~\ref{Sec:iz_def},
the main missing information is the distribution
of neutral populations across the excited states.

\subsubsection{Introduction to the collisional-radiative model}
\label{Sec:CR_model}

The problem of knowing the population distribution of excited states
is non trivial, especially for non-Maxwellian velocity distributions.
However, two limit states are well-known. The first one is at thermal
equilibrium and is characteristic of highly collisional (dense) plasmas.
In this case, the system does not radiate energy out,
which means that every photon emission is compensated with the
absorption of a photon with the same energy.
By detailed balancing one can compute the distribution
of excited states, which gives a Boltzmann distribution,
as well as a Planck distribution for the photon energy.
This regime is sometimes called \emph{local ionization equilibrium}
or \emph{collisional ionization equilibrium} \cite{dopita_2003}.
The second one has been addressed in the previous section and is
characteristic of low density plasmas ($\lesssim 10^{18}m^{-3}$).
There, the typical time between two electron impact excitations
is very long ($\sim 1/\nu_\text{Iz}\gg 10^{-6}$s at $10\eV$) compared
to the de-excitation time for an excited state of a hydrogen atom
($\sim 10^{-8}$s \cite{summers_2006}). This leads
to the overwhelming majority of the neutrals being in the
ground state. In turn, it means that virtually every
ionization is a single-step ionization, and that
some non-negligible fraction of the energy is lost
to radiations. This equilibrium is often called \emph{coronal equilibrium},
because it was believed to be applicable to the sun corona \cite{stangeby_2000}.
This regime is fundamentally based on a time scale separation
assumption between a short de-excitation time and a comparatively
long electron-neutral collision time.
Coronal equilibrium is therefore characteristic of weakly
collisional plasmas.

Usually, for tokamak plasmas, the answer lies in the middle.
Radiating out of the plasma, strong emissions in the Balmer series
are observed experimentally \cite{karhunen_2025}. Firstly, this means that
a significant amount of photons are produced by the decay of a higher ($\geq 3$)
excited state to the second one (of energy $E_2$). This indicates a significant
importance of multi-step processes. Secondly, these emissions are not absorbed
back by the plasma itself (which is called \emph{transparency}, or
\emph{optically thinness} to
the Balmer emissions), indicating a relatively low fraction of
the neutrals being in the excited state $E_2$. Conversely,
the plasma usually does not emit in the Lyman series,
\cite{karhunen_2025}. It means that the Lyman emissions are
absorbed back in the plasma
(called \emph{opacity} or \emph{optical thickness}
to the Lyman emissions). This is explained
by the fact that a large fraction of the neutrals are in the ground state.

The general problem of knowing the population distribution of neutrals
across excited states is hard to solve.
However, some tools have been developed to propose approximate solutions.
One of the most widely used models,
the Collisional-Radiative Model (CRM), see \cite{summers_2006}, assumes
a Maxwellian plasma, as well as a quasi-equilibrium,
which means that the distribution of excited states for neutrals is
governed by the density and temperature of the electron species.
The CRM then proceeds to compute the distribution of the neutral
excited states, under the hypothesis of optically thin plasma
(not taking into account any re-absorption of photons
\footnote{This is the main limitation of the CRM, see \cite{karhunen_2025}.
	A better hypothesis would be to assume opacity to Lyman, and transparency
	to all other emissions. However, the CRM has been mainly developed to
	study impurities in the edge plasma, where these hypotheses are viable.
}).
This is done by solving a linear system coming from the
balance between the de-excitation times of all energy states
of the atom, and the typical time between two excitation events,
weighted by the average energy of the exciting electron.
Of course, one needs to restrict the number of energy levels
considered, usually around $40$.
The output of the calculation is $N_k(n_e, T_e)$,
the ratio between the density of a given state $k$
and the neutral density.
Obviously, the constraint $\sum_k N_k=1$ should be fulfilled.

By means of the CRM, one can compute two relevant quantities
for the problem at hand, if we assume proximity for a non-Maxwellian plasma
to the CRM predictions. The first one is an effective reaction rate,
named $\svi^\text{CRM}$ hereafter.
Usually, reaction rates are computed as
\begin{equation}
	\svi(T_e) = \int d^3\av\; v
	\siEv \Max{e}(\av),
	\label{eq:def_sv_iz}
\end{equation}
with $\Max{e}$ a normalised ($n_e=1$) centred Maxwellian distribution
for electrons with temperature $T_e$.
The effective reaction rate is
\begin{equation}
	\svi^\text{CRM}(n_e,T_e)=\sum_k N_k(n_e,T_e) \int d^3\av\; v
	\si^{k}\left(\Eev\right)\Max{e}(\av),
	\label{eq:def_sv_iz_CRM}
\end{equation}
where $\si^k$ is the cross section for Reaction~\eqref{ionization_excited_Eq}.
The case $k=1$ is the ground state, presented in
Section~\ref{Sec:iz_ground_state}.
The second quantity of interest is
the total energy loss by radiation due to the transitions
between each of the excitation states, the so-called
\emph{energy weighted reaction rate}, denoted $\Esv_I[\eV m^3s^{-1}]$.
It reads
\begin{equation}
	\Esv_I^\text{CRM}(n_e,T_e)=\frac{1}{n_e}\sum_{j>k}
	N_j(n_e,T_e)N_k(n_e,T_e) A_{j\to k}[E_j-E_k],
	\label{eq:def_Esv_iz_CRM}
\end{equation}
in order to have $P_\text{rad}=n_en_N\Esv_I^\text{CRM}$ the radiated power density.
Here $A_{j\to k}[s^{-1}]$ is the Einstein coefficient for spontaneous emission
to decay from state $j$ to $k$.
Note that depending on the convention, this $\Esv_I^\text{CRM}$
can take into account the energy lost in order to ionize the atom,
or only the radiation loss.
To stay consistent with the operator derived in
Section~\ref{Sec:iz_ground_state} we choose the latter.
It means that the total energy loss to radiation should be
$\Esv_I^\text{CRM}+E_I\svi^\text{CRM}$.
Both $\sv_I^\text{CRM}$ and $\Esv_I^\text{CRM}$, amongst others, are
stored in the ADAS database.
We emphasize here that these quantities depend on the plasma temperature
and density (assumed to be the electron ones), which is
not the case of Eq.~\eqref{eq:def_sv_iz} (that only depends
on $T_e$). However, when taking the density to zero,
$\sv_I^\text{CRM}$ converges towards $\svi$
(given by coronal equilibrium, see the FLYCHK database
\cite{chung_2005}), as expected.
We now proceed to show how to practically use the above discussion.

\subsubsection{Application through dressed cross sections}

In order to capture the effect of excitation levels amongst neutrals
in the derivation of Section~\ref{Sec:iz_ground_state},
we need to introduce two changes.

Firstly, the calculations performed in Section~\ref{Sec:iz_ground_state}
are underestimating the probability
of a neutral being ionized by neglecting its actual energy level.
Indeed, the $\si$ and $d\sigma/dE$ used in Eq.~\eqref{eq:termf_iz_ion},
\eqref{eq:term_sd_sink}, \eqref{eq:term_sd_source_approx}
and \eqref{eq:termej} assume the incoming neutral is in ground
state. We correct this by artificially increasing the value
of $\si$ and $d\sigma/dE$ to match the collisional-radiative model given
in Section~\ref{Sec:CR_model}.
This has been introduced previously as \emph{dressed cross sections}
in \cite{tskhakaya_2023}. We define
\begin{gather}
	\begin{aligned}
		\effsi(E, n_e, T_e)
		 & = D_n(n_e, T_e)\;\si(E),
		\\
		\effdsde(\Ein, E, n_e, T_e)
		 & =D_n(n_e, T_e)\;\dsde(\Ein, E),
		\\
		D_n(n_e, T_e)
		 & = \frac{\svi^{\text{CRM}}(n_e,T_e)}{\svi(T_e)},
	\end{aligned}
	\label{eq:def_dressed_cross_sections}
\end{gather}
with $\svi$ defined in Eq.~\eqref{eq:def_sv_iz}.
Note that the multiplicative density dressing factor $D_n$ does not depend
on the velocity variable. This means we will always
be able to take it out of all integrals. This is the main advantage of this
solution, but also its main downside. Indeed, as it is a global rescaling,
electrons with less than $13.6\eV$ will not be able to ionize any neutrals,
even though, in reality, they could ionize the excited population of neutrals.
Another solution would have been to compute an effective cross section by
\begin{equation*}
	\si^\text{eff}(E,n_e,T_e)=\sum_k N_k(n_e,T_e) \si^k(E).
\end{equation*}
The drawback is that it would require to solve the
CRM at every time step to compute the $N_k$ concentrations of all
excited states rather than
relying on tabulated $\svi$, and would forbid the use of pre-computed
data. This is why this alternative solution has not been retained.
In the remaining of this work we drop the $n_e$ and $T_e$ dependencies
for the sake of readability.

Secondly, we should introduce an energy loss term due to radiation,
$\Scal_\rad$.
This energy loss is assumed to only impact the electron species.
This term should be a pure sink of energy,
with amplitude given by $n_e n_N\Esv_I^\text{CRM}$.
The complete kinetic term is complex to derive, and necessitates
the handling of several excitation states. In this work we choose
a very simplified approach, which nevertheless retains the main physical feature.
As presented in Section~\ref{Sec:CR_model}, tokamak plasmas
mostly emit radiation in the Balmer series. Moreover, as
$N_k$ decays exponentially with $k$ (as it is close to a
Boltzmann distribution, see
Section~\ref{Sec:CR_model}), the dominant
effect is seen through Balmer-$\alpha$ emissions (decaying
from energy state $E_3$ to $E_2$). In the following, we assume that
the entire radiation loss is carried through Balmer-$\alpha$ emissions.
The most probable cause of such an emission is the excitation
of an atom from the ground state to the energy state $E_3$ by
electron impact, since the ground state is the most abundant
energy state. The effect is then a net loss of energy for the
electron species of $\Delta E_{1\to 3}=|E_3-E_1|$. We note here
that $\Delta E_{1\to 3}$ is greater than the energy loss by radiation
(which is only $\Delta E_{2\to 3}$). The rest of this energy goes
into the potential energy of the population of neutrals, and
participates in increasing $\si^\eff$ as seen in
Eq.~\eqref{eq:def_dressed_cross_sections}.

Therefore, $\Scal_\rad$ should remove electrons from the
distribution function according to
the excitation cross section $\sigma_{1\to 3}$ and add them
back with $\Delta E_{1\to 3}$ less energy.
Such a cross section is given for instance by the NIST database
\cite{nist_2009}.
This leads to
\begin{multline}
	\mathcal{S}_{\text{rad},e}(\av) =
	D_E n_N
	\Bigg[
	\left[v+\dfrac{2\Delta E_{1\to 3}}{m_e v}\right]
	\sigma_{1\to 3}\left(\Eev+\Delta E_{1\to 3}\right)
	f_e\left(\frac{\av}{v}\sqrt{v^2+\frac{2\Delta E_{1\to 3}}{m_e}}\right)
	\\
	{} - v\sigma_{1\to 3}\left(\Eev\right)f_e(\av)\Bigg]
	\label{eq:energy_sink_rad}
\end{multline}
with
\begin{equation}
	D_E(n_e, T_e)=
	\frac{\Esv_I^\text{CRM}(n_e,T_e)}{\Delta E_{1\to 3} \sv_{1\to 3}(T_e)}.
	\label{eq:def_C_E}
\end{equation}
We used the notation
\begin{equation}
	\sv_{1\to 3}(T_e) = \int d^3\av\; v
	\sigma_{1\to 3}\left(\Eev\right)\Max{e}(\av).
	\label{eq:def_sigma_v_exc}
\end{equation}
Note the similarity between Eq.~\eqref{eq:def_dressed_cross_sections}
and \eqref{eq:def_C_E}, which are both cross section dressings, one in
density and one in energy.
The reason we need to include an energy dressing Eq.~\eqref{eq:def_C_E}
for the radiation sink is the fact that we only considered
Balmer-$\alpha$ emissions, thus underestimating the energy loss.
This needs to be corrected in order to match the CRM predictions:
Eq.~\eqref{eq:energy_sink_rad} leads to an energy sink
of $n_e n_N \Esv_I^\text{CRM}$ and preserves density.

In summary, the previously derived equations can be modified
to account for excited states of neutrals.
Eq.~\eqref{eq:termf_iz_ion} becomes
\begin{equation}
	\termf {i}{I} (\aw)
	= f_N(\aw) \int |\aw-\av|\effsi\left(\frac{1}{2}m_e|\aw-\av|^2\right)f_e(\av)\;d^3\av,
	\label{eq:termf_iz_ion_eff}
\end{equation}
and the final kinetic term for electrons is
\begin{equation}
	\begin{aligned}
		\termf eI (\av)=
		 & -\int d^3\aw\;|\av-\aw|\effsi(E_\av)f_N(\aw)f_e(\av)
		\\
		 & + m_e\int d^3\aw \;f_N(\aw)
		\int_1^\infty d\lsd\;
		\lsd^3|\av-\aw|^3
		\effdsde(\lsd^2E_\av,E_\av)f_e(\lsd\av).
		\\
		 & + \frac{m_e}{4\pi}
		\int d^3\aw\;f_N(\aw)
		\int_{|\au-\aw|\ge|\av-\aw|}d^3\au\;
		\frac{|\au-\aw|}{|\av-\aw|}
		\effdsde(E_\au,E_\au-E_\av-E_I) f_e(\au),
		\\
		 & + \mathcal{S}_{\text{rad},e}(\av),
	\end{aligned}
	\label{eq:termf_iz_elec_eff}
\end{equation}
with $E_\av=\frac{1}{2}m_e|\av-\aw|^2$,
$E_\au=\frac{1}{2}m_e|\au-\aw|^2$ and $\mathcal{S}_{\text{rad},e}$
given by Eq.~\eqref{eq:energy_sink_rad} to \eqref{eq:def_sigma_v_exc}.
Note that, in the second-to-last line, it is still
the ionization energy $E_I$ appearing, and not an effective
one depending on the mean excitation level of neutrals.
This is because any ionization event requires an energy
of $E_I$, no matter how many excitation steps it takes.
This is a direct consequence of the global rescaling (or dressing)
of the cross sections, without solving the CRM dynamically.

This concludes the derivation of the kinetic operator for electron impact
ionization of hydrogen-like atoms. The generalization to higher $Z$ atoms
(usually referred to as \emph{impurities}) is possible, if not trivial.
For heavier atoms, ionization is not always due to the ejection
of an electron coming from the outermost shell. If the ejected electron
comes from one of the inner shells of the atom, the resulting ion is then
left in an excited state, and can de-excite by spontaneous emission.
This first prohibits the definition of one ``ionization energy'' of the atom,
the energy lost by the electron would now be dependent on the ejection shell.
It then impacts the measurement of differential cross sections, both in
energies and in angles, potentially influencing the correctness of the hypothesis
presented in Eq.~ \eqref{eq:ass_on_dsdedodo}. Additionally, ionization of inner
shells needs to be taken into account for the energy lost as radiation. Finally,
some impurities exhibit \emph{metastable} excitation levels, \textit{i.e.}
long-lived excited states, that need to be taken into account when computing
the distribution of excited states. This is left for future work, we now turn
our attention to the other two reactions studied in this article, namely recombination
and charge exchange.

\section{Kinetic operators for recombination and charge exchange}
\label{Sec:kin_recCX}

\subsection{Radiative and three-body recombination}
When recombining, an electron goes from a free state to being trapped in the potential
well of a nucleus. The latter being a lower energy state than the first one, the reaction
needs a third body to carry out the excess energy. This role is
mainly\footnote{The third body could technically be a ion, but ion assisted atomic
	recombination is orders of magnitude less frequent than the alternatives presented here}
played by electrons
(the so-called three-body recombination) and photons (the so-called radiative,
or two-body, recombination). In the case of radiative recombination,
which is the dominant process for impurities and heavy ions, the energy is lost
in the form of radiation, and therefore is a energy sink for electrons.
Conversely, three-body recombination, which is largely dominant at low temperature
(of the order of a few $\eV$) for hydrogen-like atoms, see \cite{lumma_1997},
is an energy source for electrons, as the potential energy of the free electron
in transferred to the third body.

\begin{figure}
	\centering
	\includegraphics[width=0.3\textwidth]{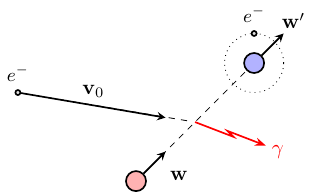}
	\caption{Radiative recombination of an ion (velocity $\aw$) with an electron
		(velocity $\av_0$) to a neutral (velocity $\aw'$). The excess energy is
		carried out by a photon $\gamma$.}
	\label{Fig:rad_rec}
\end{figure}

The radiative recombination can be seen as the inverse
process of photoionization, and can be written as
\begin{equation} \label{Rtion:rad_rec}
	H^+ + \e \rightarrow H(k) + \gamma(\nu).
\end{equation}
A sketch of this reaction is given in
Figure~\ref{Fig:rad_rec}.
Note that the outgoing recombined neutral is most generally
in a given excited state $k$.
In the context of this work, we focus on what is sometimes called
\emph{complete radiative recombination} \cite{lumma_1997},
\textit{i.e.} the case $k=1$, where the neutral radiates out
all the excess energy and goes back to ground state.
We denote complete radiative recombination by the subscript
$\text{RR}$.

\begin{figure}
	\centering
	\includegraphics[width=0.4\textwidth]{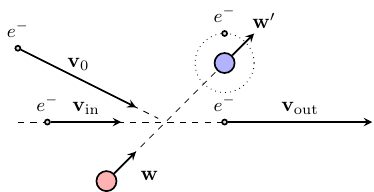}
	\caption{Three-body recombination of an ion (velocity $\aw$) with an electron
		(velocity $\av_0$) to a neutral (velocity $\aw'$). The excess energy is
		carried out by another electron incoming with velocity $\avin$
		and outgoing with a velocity $\avout > \avin$.}
	\label{Fig:3b_rec}
\end{figure}

On the other hand, three-body recombination is
the inverse process of electron impact
ionization, and can be written as
\begin{equation} \label{Rtion:3b_rec}
	H^+ + \e + \e \rightarrow H + e^-.
\end{equation}
Note that, here again, we only consider complete recombinations,
\text{i.e.} when the outgoing neutral is in the ground state.
It involves a spectator electron, to which the excess energy
is transferred.
It is depicted in Figure~\ref{Fig:3b_rec}.
For three-body recombination we use the subscript $3\text{BR}$.
For both of these reactions we neglect the momentum carried
by the electron and use the simplification $\aw=\aw'$.
As said in the introduction, in the context of this work we
will not derive the operator for three-body recombination,
and only focus on Eq.~\eqref{Rtion:rad_rec}.
Indeed, three-body recombination
cannot be treated by the model developed here, which focuses on
two-body interactions.

The effect of radiative recombination on both ion and electron
distribution functions is easily computed following the same
steps as in Section~\ref{Sec:ionization_on_ions}.
We obtain these two terms:
\begin{equation}
	\termf {i}{RR} (\aw)
	= -\int d^3\av \;|\aw-\av|
	\sigma_{\text{RR}}\left(\frac{1}{2}m_e|\aw-\av|^2\right)
	f_i(\aw)f_e(\av),
	\label{eq:termf_rec_ion}
\end{equation}
and
\begin{equation}
	\termf {e}{RR} (\av)
	= -\int d^3\aw \;|\av-\aw|
	\sigma_{\text{RR}}\left(\frac{1}{2}m_e|\av-\aw|^2\right)
	f_i(\aw)f_e(\av).
	\label{eq:termf_rec_elec}
\end{equation}
It is to be noted that Eq~\eqref{eq:termf_rec_elec} is an
energy sink for the electrons.
Per recombination event, an energy of $\Eevw$ is lost for the
electron species due to the disappearance of an electron.
The energy carried out by the photon, however, is greater, since
it contains the potential energy $E_I$ of the free electron.

The data for $\sigma_\text{RR}$ can be hard to find, especially because
most databases do not distinguish between three-body recombination, radiative
recombination and complete radiative recombination. The choice is therefore
often made to assume all recombination events to be complete radiative recombination.
This concludes the section. The last considered reaction is charge exchange, and
is the topic of the next section.

\subsection{Resonant charge exchange}
\begin{figure}
	\centering
	\includegraphics[width=0.4\textwidth]{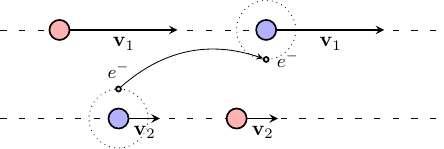}
	\caption{Charge-exchange between a slow hydrogen atom
		(velocity $\av_2$) and a fast ion (velocity $\av_1$).
		The two velocities have been depicted to be collinear, but in general
		they are not.}
	\label{Fig:cx}
\end{figure}

Charge exchange is the process by which, during
the interaction between a neutral and an ion, a bound
electron is transferred from the neutral to the incoming ion.
Although charge exchange can be viewed as the simultaneous
ionization of the neutral and recombination of the ion,
it is in fact a distinct process that is generally
much more probable than the independent occurrence
of these two events. This enhanced probability results
from the electron tunnelling between the potential wells
associated with the two nuclei.
When the energy required to ionize the neutral is equal
to the energy released upon recombination, as, for example,
in the reaction
\begin{equation*}
	H + H^+ \to H^+ + H,
\end{equation*}
no energy is radiated out (and no energy is taken from
the environment) and the process is referred to as
\emph{resonant} charge exchange. Non-resonant charge exchange
can also occur, for instance between impurities and
main ions or between excited neutrals and ions,
but is generally neglected in the study of
plasma–neutral interactions.
A sketch of what happens during charge exchange is
presented in Figure~\ref{Fig:cx}.
We denote with $\scx$ the charge exchange cross section,
depending on the relative ion and neutral velocities.
We make this dependence explicit by expressing the cross
section in terms of the impact energy
$E=\frac{1}{2}m_i|\av_1-\av_2|^2$.
Note that this definition implicitly assumes $m_i=m_N$,
otherwise the energy in the center of mass frame
would contain an additional factor of $m_i/m_N$.
For more details about the expression of $\scx$ we refer
to Appendix~\ref{Annex:cross_sections}.

There are two ways whereby charge exchange modifies the
ion distribution function at a given velocity $\av$. Either an ion with
velocity $\av$ interacts with an atom with velocity $\aw$ and exchanges
velocities, or an ion with velocity $\aw$ interacts with an atom with
velocity $\av$ and takes it.
In the case of an ion with velocity $\av$ encountering
a distribution of neutrals (at time $t$)
$f_N^t$, the probability of undergoing a charge exchange reaction
during a given time interval $\Delta t$ is
\begin{equation}
	p_i(\av)=\Delta t\int d^3\aw\; |\av-\aw|
	\scx\left(\frac{1}{2}m_i|\av-\aw|^2\right)
	f^t_N(\aw).
	\label{eq:pi_cx}
\end{equation}
The same can be said for a hydrogen atom travelling with velocity $\av$
and facing a distribution of ions $f_i^t$,
\begin{equation}
	p_N(\av)=\Delta t\int d^3\aw\; |\av-\aw|
	\scx\left(\frac{1}{2}m_i|\av-\aw|^2\right)
	f^t_i(\aw).
	\label{eq:pn_cx}
\end{equation}
After a time $\Delta t$ the distribution function has evolved according to
\begin{equation*}
	f_i^{t+\Delta t}(\av)=f_i^t(\av) - p_i(\av)f^t_i(\av) + p_N(\av)f^t_N(\av).
\end{equation*}
Inserting Eq. \eqref{eq:pi_cx} and \eqref{eq:pn_cx}, taking $f_i^t(\av)$
to the other side, dividing by $\Delta t$ and taking the limit $\Delta t\to 0$ we obtain
\begin{equation}
	\termf{i}{cx}(\av) =
	\int d^3\aw\;
	|\av-\aw| \scx\left(\frac{1}{2}m_i|\av-\aw|^2\right)
	\Big[f_i(\aw)f_N(\av)-f_i(\av)f_N(\aw)\Big].
	\label{eq:termf_cx}
\end{equation}
We remark that Eq.~\eqref{eq:termf_cx} is consistent with
Eq.~(2) in \cite{marandet_statistical_2010}, which is the same
applied to the neutral distribution function.

This concludes Section~\ref{Sec:kin_recCX}, where
we have derived the kinetic terms for
recombination, Eq.~\eqref{eq:termf_rec_ion} and \eqref{eq:termf_rec_elec},
and charge exchange, Eq.~\eqref{eq:termf_cx}.
In the next section we derive their gyrokinetic counterparts,
as well as the ionization one.

\section{Moving towards gyrokinetics}
\label{Sec:gyro_derivation}

\subsection{Preliminaries on gyrokinetics}
\label{Sec:gyro_prel}

The gyrokinetic framework is obtained from a kinetic description
by decomposing the dynamics of charged particles in a magnetic field
in a fast cyclotron gyration, a parallel advection and
slower perpendicular drifts.
By using the time scale separation between the fast gyromotion and the
(relatively) slower turbulence dynamics, one can average out the fast time
scale and end up with five-dimensional distribution
functions (instead of six). These distribution functions describe
the motion of the particle gyrocentres. For a more in-depth dive into
the derivation of gyrokinetics, we refer to \cite{brizard_2007},
\cite{frieman_1982}, \cite{garbet_2023},
\cite{tronko_2017} and \cite{tronko_2018}.

Given a spatial position $\ax$, we express the problem using
cylindrical coordinates along the magnetic field at $\ax$.
The velocity vector along the magnetic field
is $\av_\pll$, and any velocity in the transverse
plan is labelled $\av_\perp$. A given velocity can therefore
be expressed $\av=(v_\pll, v_\perp\cos\zeta, v_\perp\sin\zeta)$,
with $\zeta$ the gyrophase of the cyclotron motion.
The particle distribution function of each species is then written in
the new basis:
\begin{equation}
	f_s(\av)=f_s(v_\pll, v_\perp, \zeta),
	\quad s\in\{i,e,N\}.
\end{equation}
The guiding-centre distribution function of each species is denoted
$\of_s(v_\pll, v_\perp)$, and does not depend on the phase.
These quantities $\of_{e,i}$ are the ones evolved
by a gyrokinetic codes like \textsc{Gysela} \cite{grandgirard_2016}.
In this work, as in \cite{donnel_2021}, we neglect the spatial
effect of finite Larmor radius (this corresponds to the so-called
drift-kinetic approach) and assume the particle to be
at the location of its gyrocentre.

\begin{figure}
	\centering
	\includegraphics[width=0.5\textwidth]{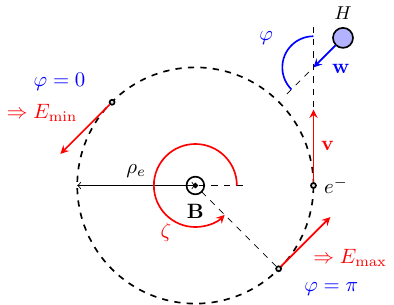}
	\caption{Impact energy for ionization along the electron gyromotion
		(angle $\zeta$), for a given incident neutral.
		In the derivation we consider the drift-kinetic limit $\rho_e\to 0$,
		yet retaining the variation of velocity direction of the electron
		along its gyromotion.}
	\label{Fig:gyromotion}
\end{figure}

However, during its gyromotion, the velocity of the particle
changes direction. This variation needs to be taken into account
when dealing with plasma-neutral interactions. Indeed, consider
an ionization reaction, depicted in Figure~\ref{Fig:gyromotion}.
Note that the following reasoning is valid for all the reactions,
not only ionization.
Depending on where the electron is on its gyromotion, the impact energy
changes. Let us call $\varphi$ the angle between $\aw$, the neutral
velocity, and $\av$ the electron velocity.
When expressing $\av$ and $\aw$ in the parallel and perpendicular
directions,
\begin{equation}
	\av=(v_\pll, v_\perp\cos\zeta, v_\perp\sin\zeta),
	\quad \aw=(w_\pll, w_\perp\cos\xi, w_\perp\sin\xi),
	\quad \varphi=|\zeta-\xi|.
	\label{eq:v_w_in_cylindrical}
\end{equation}
When $\varphi=0+2\pi n$ for $n\in\NN$,
\textit{i.e.} when $\av$ and $\aw$ are collinear pointing
in the same direction,
the impact energy is minimal ($|\av-\aw|<|\av|$).
On the other hand, when $\varphi=\pi+2\pi n$ for $n\in\NN$,
\textit{i.e.} when $\av$ and $\aw$ are collinear
with opposite directions, the impact energy is maximal ($|\av-\aw|>|\av|$).
Generally speaking, the impact energy for ionization is given by
\begin{equation}
	E^\text{iz}(\varphi)=\Eminiz+2m_e v_\perp w_\perp \sin^2(\varphi/2)
	\in \big[\Eminiz, \Emaxiz\big]
	\label{eq:impact_energy_wrt_Emin}
\end{equation}
with
\begin{equation}
	\begin{aligned}
		\Eminiz & = \frac{1}{2}m_e \Big[[v_\pll-w_\pll]^2+[v_\perp-w_\perp]^2\Big],
		\\
		\Emaxiz & = \frac{1}{2}m_e \Big[[v_\pll-w_\pll]^2+[v_\perp+w_\perp]^2\Big].
	\end{aligned}
	\label{eq:E_min_iz}
\end{equation}
For all the kinetic terms derived in Sections~\ref{Sec:kin_iz}
and \ref{Sec:kin_recCX}
we now need to compute the corresponding gyrokinetic terms by
averaging over the velocity along the gyromotion.
That is, we compute, for any reaction $\mathcal{R}$,
\begin{equation}
	\termof{i,e}{\mathcal{R}}(v_\pll, v_\perp)
	=\frac{1}{2\pi}\int_0^{2\pi} d\zeta\;
	\termf{i,e}{\mathcal{R}}\left(v_\pll, v_\perp\cos\zeta, v_\perp\sin\zeta\right).
\end{equation}
This is done in Appendix~\ref{Annex:gyro_details}
and the results are presented in Section~\ref{Sec:gyro_der}.
The integrals are presented over the angle $\theta=\varphi/2$ due to symmetries
of the problem.
Moreover, some of the terms only depend on the
difference between $v_\pll$ and $w_\pll$, that we denote
$\deltavpll=|v_\pll-w_\pll|$.

\subsection{Presentation of the gyrokinetic terms}
\label{Sec:gyro_der}

In this section we present the gyrokinetic terms
corresponding to the kinetic ones derived in
Sections~\ref{Sec:kin_iz} and \ref{Sec:kin_recCX}.
The details of the computations are shown in
Appendix~\ref{Annex:gyro_details}.

For the charge exchange reaction, Eq.~\eqref{eq:termf_cx} becomes
\begin{multline}
	\termof{i}{\cx}(v_\pll, v_\perp)
	= \int dw_\pll \int dw_\perp\;w_\perp
	\Gcx\big(|v_\pll-w_\pll|, v_\perp, w_\perp\big)
	\Big[\of_N(v_\pll, v_\perp) \of_i(w_\pll, w_\perp)
		\\
		- \of_{i}(v_\pll, v_\perp) \of_N(w_\pll, w_\perp)\Big],
	\label{eq:gyro_term_cx}
\end{multline}
with
\begin{gather}
	\begin{aligned}
		\Gcx(\deltavpll ,v_\perp, w_\perp)
		 & = 4\int_0^{\pi/2} d\theta\;
		\scx\Big(E^\cx(\theta)\Big)
		\sqrt{2E^\cx(\theta)/m_i},
		\\
		E^\cx(\theta)
		 & =\Emincx+2m_i v_\perp w_\perp \sin^2\theta,
		\qquad
		\Emincx = \frac{1}{2}m_i
		\Big[\deltavpll^2 + |v_\perp - w_\perp|^2\Big].
	\end{aligned}
	\label{eq:def_G_CX}
\end{gather}
We denoted $\Gcx$ the velocity gyroaverage of the quantity
$|\av-\aw|\scx\left(\frac{1}{2}m_i|\av-\aw|^2\right)$.

Note that in Eq.~\eqref{eq:gyro_term_cx} the neutral distribution
function only appears under the form $\of_N$, which is defined by
\begin{equation}
	\of_N(v_\pll, v_\perp)=\frac{1}{2\pi}\int_0^{2\pi}d\zeta\;
	f_N(v_\pll, v_\perp, \zeta).
	\label{eq:def_average_fN}
\end{equation}
This definition naturally arises in the calculation, and
is not the result of any additional simplification.

Concerning ionization, we introduce the notation
\begin{equation*}
	\left\langle\mathcal{A}(w_\pll, w_\perp)\right\rangle_{\of_N}
	= \int dw_\pll \int dw_\perp\; w_\perp
	\of_N(w_\pll, w_\perp) \mathcal{A}(w_\pll, w_\perp)
\end{equation*}
for any function $\mathcal{A}$.
We find for the ion term, Eq.~\eqref{eq:termf_iz_ion},
and for the incident electron term,
\begin{align}
	\termof{i}{I}(v_\pll, v_\perp)
	 & =\of_N(v_\pll, v_\perp) \int dw_\perp\;w_\perp
	\int dw_\pll\;
	\Gi(|v_\pll-w_\pll|, v_\perp, w_\perp)
	\of_e(w_\pll, w_\perp)
	\label{eq:gyro_term_iz_ion}
	\\
	\termofE(v_\pll, v_\perp)
	 & =\of_e(v_\pll, v_\perp)
	\left\langle
	\Gi(|v_\pll-w_\pll|, v_\perp, w_\perp)
	\right\rangle_{\of_N}
	\label{eq:gyro_term_in}
\end{align}
with
\begin{gather}
	\begin{aligned}
		\Gi(\deltavpll,v_\perp, w_\perp)
		 & = 4\int_0^{\pi/2} d\theta\;
		\effsi\Big(E^\text{iz}(\theta)\Big)
		\sqrt{2E^\text{iz}(\theta)/m_e},
		\\
		E^\text{iz}(\theta)
		 & =\Eminiz+2m_ev_\perp w_\perp \sin^2\theta.
	\end{aligned}
	\label{eq:def_Gi}
\end{gather}
In the new notations, $\Eminiz$ now reads
\begin{equation}
	\Eminiz= \frac{1}{2}m_e\Big[\deltavpll^2 + |v_\perp - w_\perp|^2\Big].
	\label{eq:def_Eminiz_deltavpll}
\end{equation}
Note that $\Gi(\deltavpll, v_\perp, w_\perp)=
	\Gi(\deltavpll, w_\perp, v_\perp)$.

For the scattered and the ejected electron terms,
along with Eq.~\eqref{eq:v_w_in_cylindrical}, we add the
decomposition
\begin{equation}
	\au=(u_\pll, u_\perp\cos\eta, u_\perp\sin\eta).
	\label{eq:u_in_cylindrical}
\end{equation}
The result is
\begin{equation}
	\termofEe(v_\pll, v_\perp)
	= \int_{0}^\infty d\deltau\; \deltau^3
	\of_e\left(\frac{v_\pll\deltau}{v}, \frac{v_\perp\deltau}{v}\right)
	\left\langle
	G_\sc\big(|v_\pll-w_\pll|,v_\perp, w_\perp, \deltau\big)
	\right\rangle_{\of_N},
	\label{eq:gyro_term_sc}
\end{equation}
with
\begin{gather}
	\begin{aligned}
		G_\sc(\deltavpll, v_\perp, w_\perp,\deltau)
		= 4m_e \int_0^{\pi/2} d\theta\;
		\left[\frac{2E^\text{iz}(\theta)}{m_e}\right]^{-1/2}
		\effdsde\Big(\frac{1}{2}m_e \deltau^2,
		E^\text{iz}(\theta)\Big),
	\end{aligned}
	\label{eq:def_Gsc}
\end{gather}
and
\begin{equation}
	\termofej(v_\pll, v_\perp)
	\simeq \int_{0}^\infty d\deltau\; \deltau^3
	\left\langle
	G_\ej\big(|v_\pll-w_\pll|, v_\perp, w_\perp, \deltau\big)
	\right\rangle_{\of_N}
	\int_{-\pi/2}^{\pi/2} d\xi \;
	\cos(\xi)
	\of_e\big(\deltau \sin(\xi),\deltau \cos(\xi)\big),
	\label{eq:gyro_termej}
\end{equation}
with
\begin{equation}
	G_\ej(\deltavpll, v_\perp, w_\perp,\deltau)
	= 2m_e \int_0^{\pi/2} d\theta\;
	\left[\frac{2E^\text{iz}(\theta)}{m_e}\right]^{-1/2}
	\effdsde\left(\frac{1}{2}m_e \deltau^2,
	\frac{1}{2}m_e \deltau^2-
	E^\text{iz}(\theta) - E_I\right),
	\label{eq:def_Gej}
\end{equation}
where $E_\text{iz}(\theta)$ is defined in Eq.~\eqref{eq:def_Gi},
$\Eminiz$ in Eq.~\eqref{eq:def_Eminiz_deltavpll} and
$v=\sqrt{v_\pll^2+v_\perp^2}$.
Note that the $\simeq$ sign refers to the fact that we neglected
the neutral velocity contribution appearing inside $\of_e$, as in
Eq.~\eqref{eq:term_sd_source_approx}. As in Eq.~\eqref{eq:term_sd_source_approx},
this introduces an error that should be numerically corrected \textit{a posteriori}.

The last part of Eq.~\eqref{eq:termf_iz_elec_eff} is the
radiation sink.
This term is not in integral form, and does not
depend on the neutral distribution function. The result
is therefore direct:
\begin{multline}
	\bar{\mathcal{S}}_{\text{rad},e}(v_\pll, v_\perp) =
	C_E n_N
	\Bigg[
		\left[v+\dfrac{2\Delta E_{1\to 3}}{m_e v}\right]
		\sigma_{1\to 3}\left(\Eev+\Delta E_{1\to 3}\right)
		\\
		\of_e\left(\frac{v_\pll}{v}\sqrt{v^2+\frac{2\Delta E_{1\to 3}}{m_e}}
		,\frac{v_\perp}{v}\sqrt{v^2+\frac{2\Delta E_{1\to 3}}{m_e}}\right)
		{} - v\sigma_{1\to 3}\left(\Eev\right)
		\of_e(v_\pll,v_\perp)\Bigg].
	\label{eq:gyro_Srad}
\end{multline}

Lastly, we present the gyrokinetic counterparts
of recombination, Eq.~\eqref{eq:termf_rec_ion} and \eqref{eq:termf_rec_elec}.
These two terms are very similar (like Eq.~\eqref{eq:gyro_term_iz_ion}
and \eqref{eq:gyro_term_in} were similar), and share common notations.
We obtain
\begin{align}
	\termof{i}{\text{RR}}(v_\pll, v_\perp)
	 & =-\of_i(v_\pll, v_\perp) \int dw_\perp\;w_\perp
	\int dw_\pll\; \of_e(w_\pll, w_\perp)
	\Gr\big(|v_\pll-w_\pll|, v_\perp, w_\perp\big),
	\\
	\termof{e}{\text{RR}}(v_\pll, v_\perp)
	 & =-\of_e(v_\pll, v_\perp) \int dw_\perp\;w_\perp
	\int dw_\pll\; \of_i(w_\pll, w_\perp)
	\Gr\big(|v_\pll-w_\pll|, v_\perp, w_\perp\big).
\end{align}
This time, we have
\begin{gather}
	\begin{aligned}
		\Gr(\deltavpll,v_\perp, w_\perp)
		 & = 4\int_0^{\pi/2} d\theta\;
		\srr\Big(E^\text{RR}(\theta)\Big)
		\sqrt{2E^\text{RR}(\theta)/m_e},
		\\
		E^\text{RR}(\theta)
		 & =\Eminr+2m_e v_\perp w_\perp \sin^2\theta,
		\qquad
		\Eminr = \frac{1}{2}m_e
		\Big[\deltavpll^2 + |v_\perp - w_\perp|^2\Big].
	\end{aligned}
	\label{eq:def_Gr}
\end{gather}
the recombination counterpart of Eq.~\eqref{eq:def_Gi}.
As for ionization, we have
$\Gr(\deltavpll, v_\perp, w_\perp)=\Gr(\deltavpll, w_\perp, v_\perp)$.

This concludes the derivation of the kinetic and gyrokinetic operators
for all three reactions. In the next section we compare these newly derived
terms to their fluid equivalents.

\section{Comparison with a fluid approach}
\label{Sec:comp_fluid}

When speaking about fluid descriptions of a plasma,
one usually refers to the first three fluid moments of
a distribution function: the density $n$,
the momentum $\Gamma$ and the energy $\mathcal E$.
Sometimes, the fluid velocity $\mathbf{u}=\Gamma/n$ is used
instead of the momentum.
A fluid approach of plasma-neutrals interaction therefore provides
sources and sinks for these three moments, for each of the species
involved. We denote them $S_{n,s}$, $\mathbf{S}_{\Gamma,s}$
and $S_{\EE,s}$, with $s$ the species.
The goal of this section is twofold. First, we check the coherence
between the terms derived in Sections~\ref{Sec:kin_iz} to
\ref{Sec:gyro_derivation} and a fluid approach under the right
set of hypotheses. Then we compare the two models more precisely,
both in terms of their moments and their effects in velocity space.

\subsection{Plasma-neutral interactions with fluid hypotheses}
\label{Sec:fluid_hyp}

Fluid models rely on the assumption that
the plasma is collisional enough the have a distribution function
which can be approximated by a Maxwellian. In the context of electron impact
ionization and recombination, this allows us to define averaged quantities,
where by averaged we mean integrated next to a Maxwellian distribution
function. Take for instance Eq.~\eqref{eq:term_sd_sink}.
If we assume $f_e$ to be a centred Maxwellian,
noted $f_e(\av) = n_e\Max{e}(\av)$, and $f_N$ to be a shifted one,
noted $f_N(\aw) = n_N\Max{N}(\aw)$, the electron density variation becomes
\begin{equation*}
	\int d^3\av\; \termE(\av)
	= n_e n_N \int d^3\av \int d^3\aw\; |\av-\aw|
	\siEvw \Max{N}(\aw)\Max{e}(\av).
\end{equation*}
It is therefore natural to define a new quantity,
the so-called \emph{reaction rate coefficient} for ionization,
denoted $\svi$, by
\begin{equation}
	\svi(T_e,T_N) = \int d^3\av \int d^3\aw\; |\av-\aw|
	\siEvw \Max{N}(\aw)\Max{e}(\av).
	\label{eq:def_full_sv_iz}
\end{equation}
Eq.~\eqref{eq:def_full_sv_iz} can moreover by simplified
by only keeping the leading order. Indeed, if $T_N$ is similar to $T_e$,
due to the mass difference between electrons and neutrals, the typical
velocity for neutrals will be negligible compared to the typical velocity
for electrons. The fact that $T_N$ is usually smaller than $T_e$ only
reinforces this difference. Usually, see for instance \cite{reiter_2000}
or \cite{summers_2006},
$\aw$ is neglected in Eq.~\eqref{eq:def_full_sv_iz},
\textit{i.e.} $f_N$ is taken to be $\delta_\mathbf{0}$, a Dirac distribution,
which yields Eq.~\eqref{eq:def_sv_iz}.
The same can be said for recombination, for which we define
\begin{equation}
	\svr(T_e) = \int d^3\av\; |\av|
	\srrEv \Max{e}(\av).
	\label{eq:def_sv_rec}
\end{equation}
This data can be found for instance in the AMJUEL database \cite{reiter_2000},
see H.2.17 for ionization, which is for instance used by EIRENE,
and in the ADAS database, taking into account corrections
\textit{via} the CRM \cite{summers_2006}.
The data for recombination does not usually distinguish
between three-body recombination and radiative recombination,
and includes both.
Note that, in the following, and for virtually all fluid models for
plasma-neutrals interaction, the electron distribution function is
assumed to be a centred Maxwellian.
No such assumption is made for ions or neutrals.
Indeed, the ion distribution function is assumed
to be a shifted Maxwellian, $f_i(\av) = n_i\Max{i}(\av)$

For recombination, the most widely used fluid source terms
\cite{eder_2025, korving_2023, ogier_collin_2026, van_uytven_2022}
are the ones presented in \cite{militello_2022}
\begin{equation}
	\left\lbrace
	\begin{array}{rl}
		S_{n,s}^\text{RR}
		 & = -n_en_i\svr
		\\
		\mathbf{S}_{\Gamma,s}^\text{RR}
		 & = - n_e n_i \mathbf{u}_s \svr
		\\
		S_{\EE,s}^\text{RR}
		 & = -\EE_s n_e n_i \svr
	\end{array}
	\right.
	\text{ for } s\in \{e,i\}.
	\label{eq:rec_fluid_terms}
\end{equation}
We can show that the first three moments of
Eq.~\eqref{eq:termf_rec_ion} and \eqref{eq:termf_rec_elec} indeed
reduce to Eq.~\eqref{eq:rec_fluid_terms} under the fluid hypotheses
detailed above.

For ionization, the fluid source terms for the ion species
are straightforward, and can be found in \cite{militello_2022}.
This is not so clear for the electron species, at least for the
momentum and energy sinks. For the energy sink, the matter has been
discussed at length in Section~\ref{Sec:iz_CRM}, and should be different
for neutrals in the ground state and when considering excited states.
The fluid terms mentioned are standard ones and are used for instance in
\cite{eder_2025, korving_2023, ogier_collin_2026,
	rensink_1998, van_uytven_2022}.
For the momentum sink, it does not seem to exist any convention
on the question, especially since most of the fluid codes
do not solve for the electron momentum.
One of the proposed terms \cite{ogier_collin_2026} enforces momentum
conservation for the plasma.
This leads to the following terms:
\begin{equation}
	\left\lbrace
	\begin{array}{rll}
		S_{n,s}^I
		 & = n_en_i\svi,
		 & \text{ for } s\in \{e,i\}
		\\
		\mathbf{S}_{\Gamma,e}^I
		 & = -\frac{m_i}{m_e}\mathbf{u}_N n_e n_i \svi,
		\\
		\mathbf{S}_{\Gamma,i}^I
		 & = \mathbf{u}_N n_e n_i \svi,
		\\
		S_{\EE,e}^I
		 & = -E_I n_e n_i \svi
		 & \text{[ground state]}
		\\
		 & = - n_e n_i \Esv_I^\text{CRM}
		 & \text{[CRM]}
		\\
		S_{\EE,i}^I
		 & = -\EE_N n_e n_i \svi.
	\end{array}
	\right.
	\label{eq:iz_fluid_terms}
\end{equation}
We can show that the density and energy moments of
Eq.~\eqref{eq:termf_iz_ion_eff} and \eqref{eq:termf_iz_elec_eff}
reduce to Eq.~\eqref{eq:iz_fluid_terms} when using fluid hypotheses.

For charge-exchange, however , the simplification used to reduce
Eq.~\eqref{eq:def_full_sv_iz} to Eq.~\eqref{eq:def_sv_iz} is not valid anymore.
Indeed, the ion and neutral velocity are similar, and we cannot neglect one or the other.
A widely used approximation in fluid approaches is to linearize the momentum exchange,
\textit{i.e.} use a \emph{momentum linearized CX rate coefficient},
with underlying hypothesis that the momentum exchange only
depends on the difference between the ion fluid velocity
and the neutral one. This, in turn, means that $v\scx(v)$
is a constant function of $v$, see Appendix~\ref{Annex:cross_sections}
for more details.
We therefore define $\svcx$ to be that constant.
Provided that $f_i$ is a Maxwellian too,
the momentum and energy sources due to charge-exchange
have a very simple linear form that can be found in \cite{militello_2022}.
We give the formula for density, momentum and energy sources for ions,
\begin{equation}
	\left\lbrace
	\begin{array}{rl}
		S_{n,i}^\text{cx}
		 & = 0,
		\\
		\mathbf{S}_{\Gamma,i}^\text{cx}
		 & =  n_i\Gamma_N\svcx - n_N\Gamma_i\svcx,
		\\
		S_{\EE,i}^\text{cx}
		 & =  n_in_N \EE_n\svcx
		- n_Nn_i \EE_i\svcx.
	\end{array}
	\right.
	\label{eq:CX_sources_fluid}
\end{equation}
We recover Eq.~\eqref{eq:CX_sources_fluid} when
inserting the linearity and Maxwellianity assumptions
in Eq.~\eqref{eq:termf_cx} and project onto the first moments.
This model is widely used, see
\cite{eder_2025, korving_2023, ogier_collin_2026,
	rensink_1998, van_uytven_2022, wersal_first-principles_2015}.
The assumption of $v\scx(v)$ being constant is however very far from reality,
as shown in \cite{helander_1994}, and further discussed
in Appendix~\ref{Annex:cross_sections}.
This is why some codes like EIRENE \cite{reiter_2005}, and
models like \cite{helander_1994}, only assume the Maxwellianity
of the plasma, without linearizing.

In order to compare these fluid source terms to the kinetic terms derived
earlier, one needs to reconstruct a source term in velocity space from
the fluid moments presented in Eq.~\eqref{eq:rec_fluid_terms}, \eqref{eq:iz_fluid_terms}
and \eqref{eq:CX_sources_fluid}. To be coherent with fluid assumptions,
we assume the distribution for the electrons and the ions to always be a Maxwellian.
This means that the source term to go from time $t$ to time $t+\Delta t$ is the
difference between two Maxwellian distributions, namely
\begin{equation*}
	\left(\frac{df_s}{d t}\right)_{\text{fluid}}=
	\Max{s}(n_s + \Delta t S_{n,s}, \Gamma_s + \Delta t S_{\Gamma, s},
	\EE_s + \Delta t S_{\EE, s})
	-\Max{s}(n_s, \Gamma_s, \EE_s).
\end{equation*}
In the next section, these will be referred as ``fluid terms'' in velocity space,
inferred from our kinetic approach.

\subsection{Results}
\label{Sec:results}

\begin{figure}[ht]
	\centering
	\includegraphics[width=0.9\textwidth]{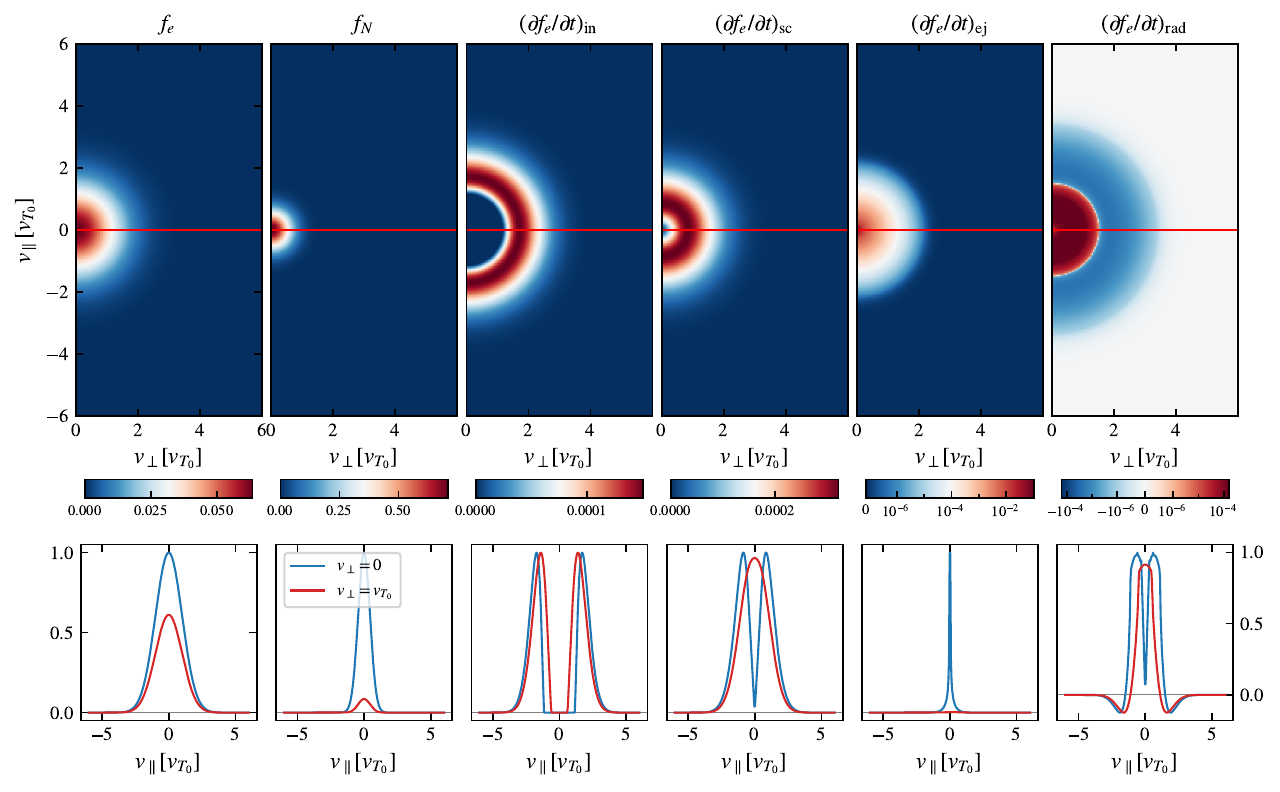}
	\caption{Eq. \eqref{eq:gyro_term_in}, \eqref{eq:gyro_term_sc},
	\eqref{eq:gyro_termej} and \eqref{eq:gyro_Srad}
	for two centred Maxwellian distributions $f_e$ and $f_N$.
	We used $T_N=4\eV$ and $T_e=20\eV$.
	The lower plots are slices at $v_\pll=0$ and $v_\pll=v_{T_0}$,
	normalised to unity.}
	\label{Fig:all_kin_terms_centred}
\end{figure}

\begin{figure}[ht]
	\centering
	\includegraphics[width=0.9\textwidth]{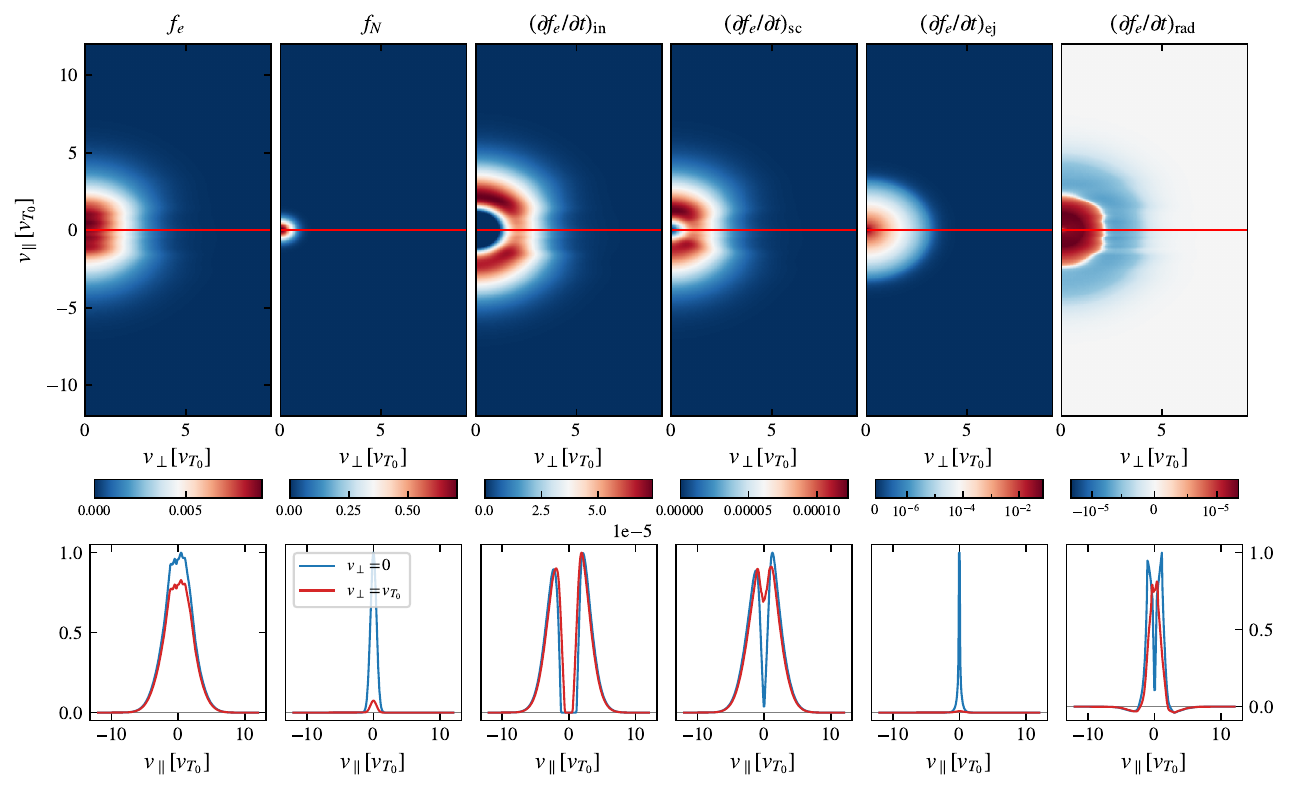}
	\caption{
		Similar to Fig.~\ref{Fig:all_kin_terms_centred}
		for a centred Maxwellian $f_N$ with temperature $T_N=4\eV$,
		and $f_e$ coming from a \textsc{Gysela} simulation.
		$f_e$ has been taken at a given space location $(r, \theta, \varphi)$.}
	\label{Fig:all_kin_terms_gys}
\end{figure}

In this section, we compute the sink and source terms due to
ionization and charge exchange both \textit{via} the kinetic description
introduced in Section~\ref{Sec:gyro_der} and the fluid one
presented in Section~\ref{Sec:fluid_hyp}. The comparison is made in
velocity space $(v_\pll, v_\perp)$, both dimensions normalized to
$v_{T_0,s}$, with $T_0$ chosen to be $20\ \eV$. We therefore plot
electron distribution functions and electron source/sink terms
normalised to $v_{T_0,e}=\sqrt{T_0/m_e}$ and the ion
distribution functions and source/sink terms to $v_{T_0,i}=\sqrt{T_0/m_i}$.
When the context is clear we only write $v_{T_0}$.
In order to compute these terms we used the cross sections
and differential cross sections presented in Appendix~\ref{Annex:cross_sections}.
We begin by presenting the ionization gyrokinetic terms for electrons,
plotted in Fig.~\ref{Fig:all_kin_terms_centred} and \ref{Fig:all_kin_terms_gys}.

Fig.~\ref{Fig:all_kin_terms_centred} shows the terms defined by
Eq. \eqref{eq:gyro_term_in}, \eqref{eq:gyro_term_sc}, \eqref{eq:gyro_termej}
and \eqref{eq:gyro_Srad}, taking $f_e$ as a centred Maxwellian with
temperature $T_e=20\eV$ and $f_N$ as a centred Maxwellian with temperature $T_N=4\eV$.
These neutrals would represent the cold population emitted by slow recycling,
coming from molecule dissociation, before thermalizing on the ions via charge exchange.
Due to the isotropy of both $f_e$ and $f_N$, all four terms are isotropic.
The incident electron term exhibits an annular structure around the origin,
with a maximum resulting from the competition between the
increasing collision frequency $n_Nv\si$ and the decreasing electron
distribution function at higher velocities.
The scattered electron term is very similar, due to the neglect
of any deflection, Eq.~\eqref{eq:ass_on_dsdedodo}.
The annular structure is, however, shifted towards the origin
compared to $\inlinetermE$, due to the energy loss of the incident electron.
The most notable feature is a pronounced peaking of $\inlinetermej$ around
zero, ejecting electrons with characteristic velocity several
orders of magnitude lower than the electron thermal speed (in the case
shown in Fig.~\ref{Fig:all_kin_terms_centred}, the typical width of
$\inlinetermej$ is less than five percent of the electron thermal velocity).
Finally, $\mathcal{S}_\rad$ takes both positive and negative values
in velocity space. It removes electrons from the high-velocity region and adds
them to the low-velocity region, so its net effect is to remove energy.

Fig.~\ref{Fig:all_kin_terms_gys} shows the same four kinetic terms,
this time using a more realistic distribution function for electrons.
This distribution function $f_e$ has been taken from a TEM-dominated
\textsc{Gysela} simulation, and exhibits non-Maxwellian features,
especially at low electron parallel velocities. These localised
features in velocity space heavily influence the shape of Eq.~\eqref{eq:gyro_term_in}
and \eqref{eq:gyro_term_sc} (third and fourth plots of Fig.~\ref{Fig:all_kin_terms_gys},
even though they are smoothed out \textit{via} the convolution with the neutrals
distribution function. The ejection term, however, remains similar, due to its
high peaking around the origin, making it less sensitive to variations in $f_e$.
The crude assumptions behind Eq.~\eqref{eq:gyro_Srad}, namely the fact that the neutral
velocity is neglected, make the radiation term inherit the structures of $f_e$,
not having been convolved against $f_N$.

\begin{figure}[ht]
	\centering
	\includegraphics[width=0.8\textwidth]{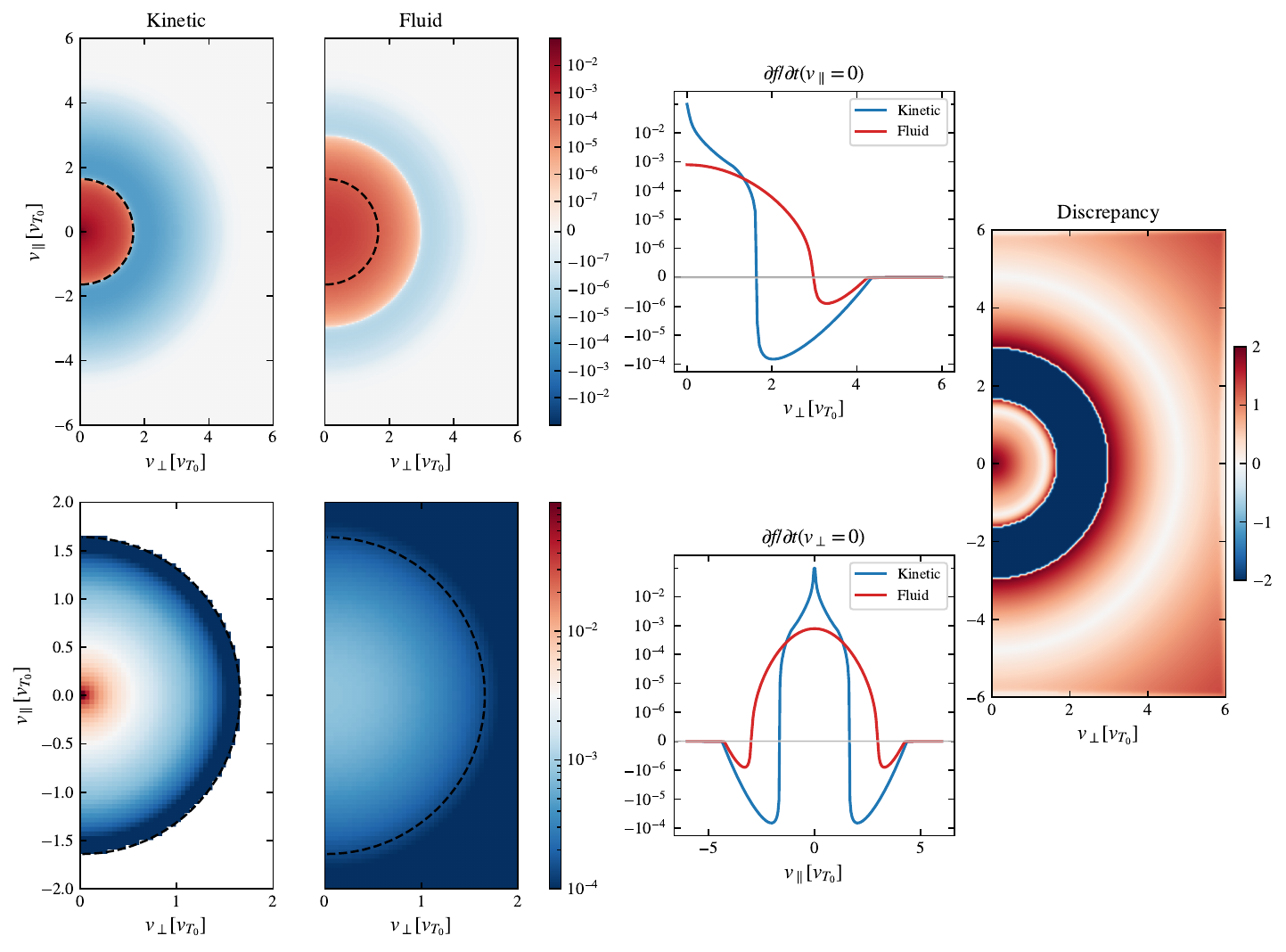}
	\caption{Comparison between the fluid and the ionization kinetic source terms for
		$T_e=20\eV$, $T_N=4\eV$. Both distribution functions are centred Maxwellians.
		The upper left two plots show the kinetic and the fluid terms for the electron species.
		The lower left two plots show the same quantities in a truncated logarithmic scale.
		The two plots in the middle show slices of both terms at $v_\pll=0$ and $v_\perp=0$.
		The dashed black lines indicate the region where the kinetic term vanishes.
		The right plot shows the relative discrepancy between the two.
		It is negative when the terms have opposite signs. }
	\label{Fig:comp_fluid_kin_20}
\end{figure}

We can now compare the resulting kinetic term with the fluid
term computed with Eq.~\eqref{eq:iz_fluid_terms}. This is presented
in Fig.~\ref{Fig:comp_fluid_kin_20}.
We choose to use only Maxwellian distribution functions
for both $f_e$ and $f_N$ to focus on the differences created
by the kinetic handling of the reaction.
We moreover show the discrepancy between the kinetic and the fluid terms:
$2|\text{fluid}-\text{kin}|/(|\text{fluid}|+|\text{kin}|)$,
multiplied by $\mathrm{sign}(\text{fluid} \times \text{kin})$, which is
positive if both terms have the same sign, and negative otherwise.

\begin{figure}[ht]
	\centering
	\includegraphics[width=0.8\textwidth]{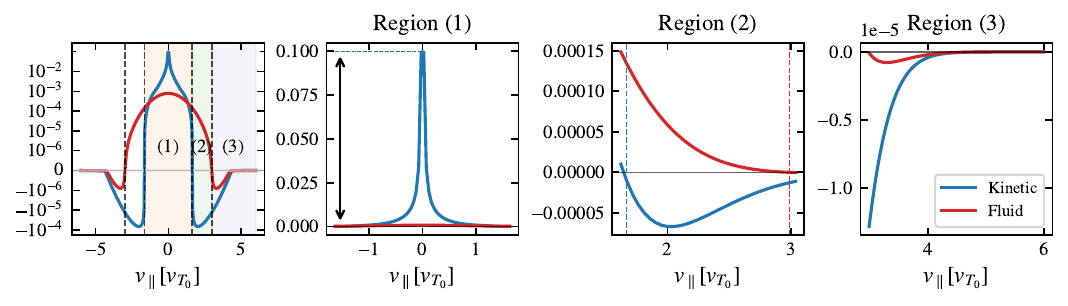}
	\caption{Splitting of the parallel slice presented in Fig.~\ref{Fig:comp_fluid_kin_20}
		into three regions of interest. The kinetic and fluid terms are plotted
		on each of the three regions. Region~(1) is defined by both terms being
		positive, region~(3) by both terms being negative and region~(2) is where
		their signs differ.}
	\label{Fig:comp_vpll}
\end{figure}

As shown in Fig.~\ref{Fig:comp_fluid_kin_20}, the kinetic and fluid terms
exhibit similar structures in velocity space, but the boundaries separating
their positive and negative regions do not coincide.
This is better seen in the one-dimensional slices.
It creates an annular region in velocity space in which the sign of both terms differ.
This region can be seen in the discrepancy plot.
In this region, the fluid term is populating the distribution function even though the
kinetic term is removing electrons.
The second thing the fluid term does not capture is the strong peaking around the origin
which results in the kinetic term from the ejection of cold electrons.
Around the origin the terms differ by several orders of magnitude.
To further investigate these discrepancies, we now focus on the slice
along the parallel velocity, shown in Fig.~\ref{Fig:comp_vpll}.

\begin{figure}[ht]
	\centering
	\includegraphics[width=0.7\textwidth]{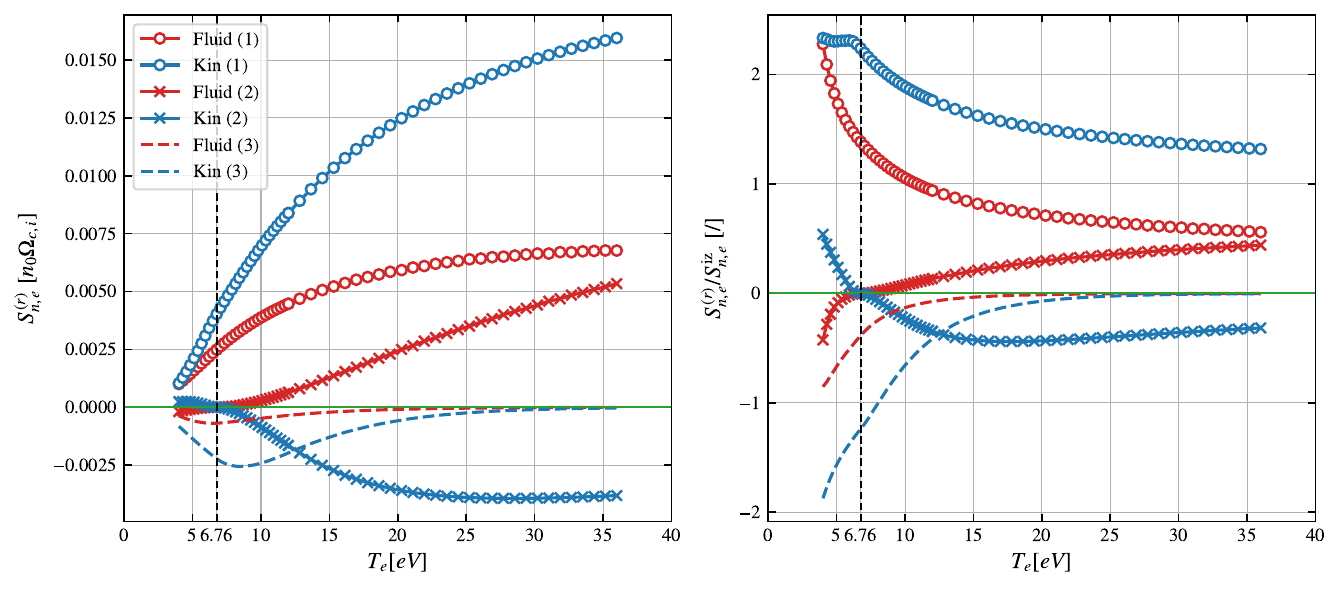}
	\caption{Contribution to the density evolution due to ionization,
		for the fluid and the kinetic terms,
		on regions~(1), (2) and (3) of velocity space.
		The left plot is the absolute value, and the right plot is relative
		to the overall density source.}
	\label{Fig:ratio}
\end{figure}

\begin{figure}[ht]
	\centering
	\includegraphics[width=0.6\textwidth]{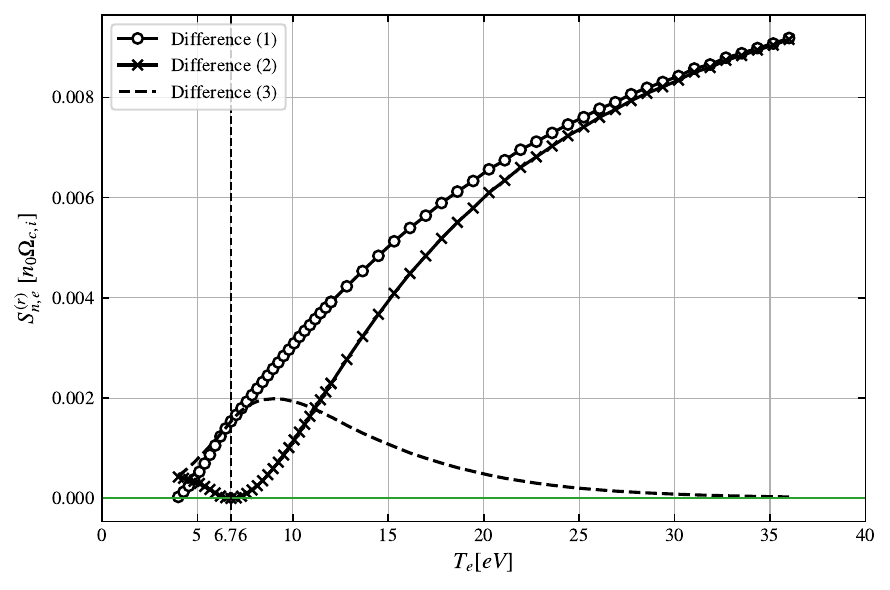}
	\caption{Absolute discrepancy between the kinetic and the fluid source term for
		ionization, on regions~(1), (2) and (3) of the velocity space.}
	\label{Fig:only_discr}
\end{figure}

Fig.~\ref{Fig:comp_vpll} identifies three regions of interest, presented
on the one-dimensional slice along $v_\pll$. The first one is the ejection region,
defined by the fact that both terms are positive. This region~(1) is marked by a large
discrepancy between the terms, as well as by a strong peaking of the kinetic term.
Region~(2) is the region for which the sign of both terms differs. In this particular
example ($T_e=20\eV, T_N=4\eV$) the fluid term is positive whereas the kinetic term is negative:
the electron sink at intermediate velocities in the kinetic approach is an
electron source in the fluid fluid approach.
Finally, we define region~(3) as the location in which both terms are negative.
In this region both terms behave in a similar way.
To better quantify these discrepancies, we compute the integral of each term on
each of these regions, corresponding to a density source or sink.
This is shown in Fig.~\ref{Fig:ratio}, and the difference between the fluid contribution and
the kinetic contribution is shown in Fig.~\ref{Fig:only_discr}.

Fig.~\ref{Fig:ratio} shows the density sinks and sources for both terms computed
by integrating over regions~(1), (2) and (3) (see Fig.~\ref{Fig:comp_vpll}).
The quantities are plotted as a function of the electron temperature.
The electron distribution function is taken
to be a centred Maxwellian distribution function with varying temperature
$T_e$. We see that the kinetic source of cold electrons (region~(1)) strongly
increases with temperature, whereas the fluid equivalent saturates starting
from $T_e\sim 25\eV$. The difference between the two therefore increases
rapidly. This difference is shown in Fig.~\ref{Fig:only_discr}.
For the region~(2) terms, we can see a different behaviour.
The difference between the fluid and kinetic terms grows with
the electron temperature. At low temperature ($T_e \sim E_I/2$),
however, we can see both terms intersect at the origin and change sign.
This means that at temperatures lower than $E_I/2$,
the fluid term is negative in region~(2), and the kinetic term is positive.
The absolute difference between them is plotted again in Fig.~\ref{Fig:only_discr},
where it rises to finally being similar to region~(1) for temperatures
higher than $30\eV$.

The right plot of Fig.~\ref{Fig:ratio} shows the kinetic and fluid contributions,
normalised to the total density source. We recall here that the density source
is the same for both terms, by construction (see Eq.~\eqref{eq:def_dressed_cross_sections}).
This means that the normalised contributions for the fluid term
and the kinetic terms both sum up to unity.
We can see that the contribution from region~(3) rapidly goes to zero as
the temperature increases, the discrepancy between the terms being dominated
by the ejection and the difference of population around the thermal velocity.
This is shown in Fig.~\ref{Fig:only_discr}.
This concludes our study of ionization.

\begin{figure}[ht]
	\centering
	\includegraphics[width=0.7\textwidth]{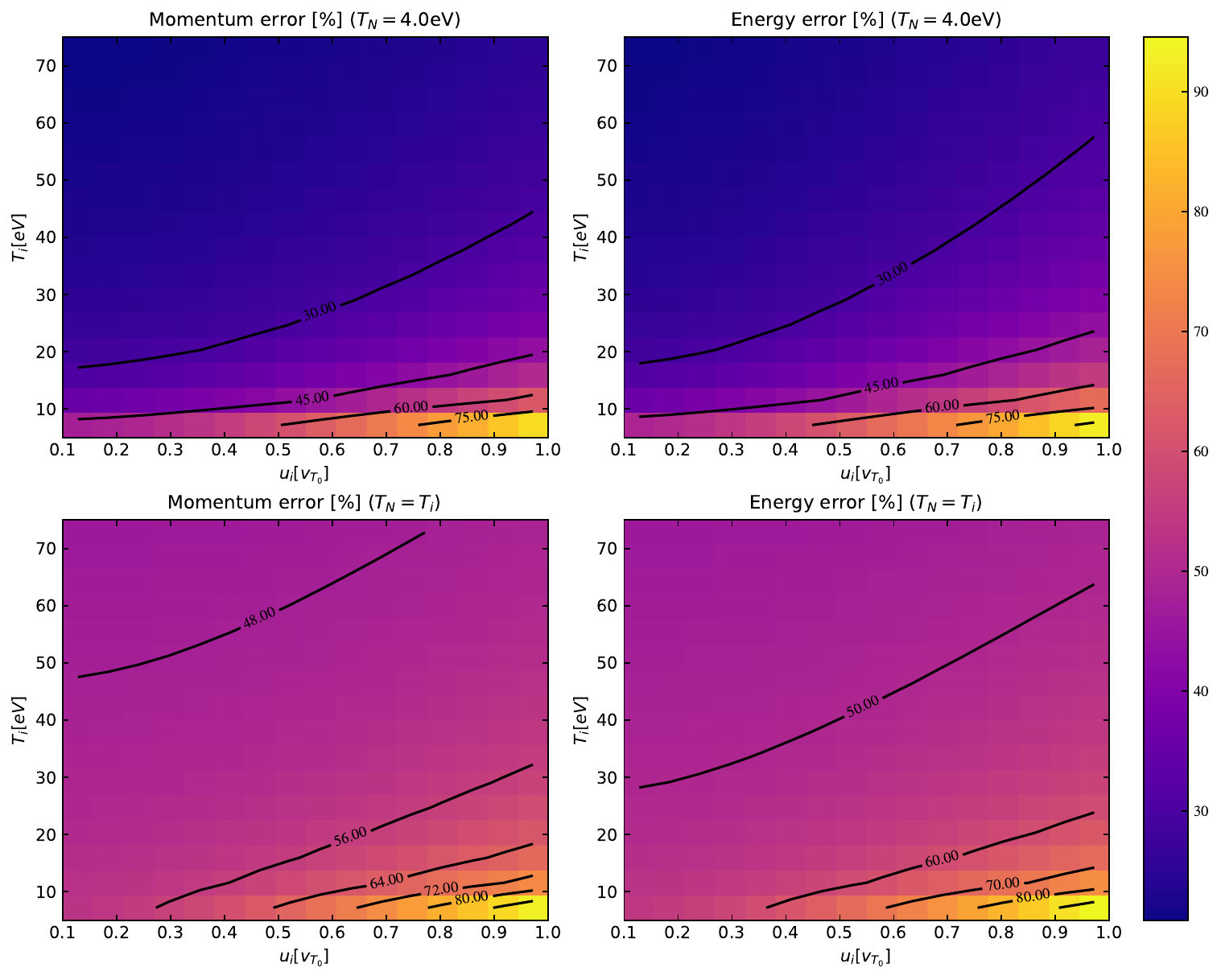}
	\caption{Charge exchange discrepancy for momentum and energy exchange,
		both for hot and cold neutrals. The discrepancy is shown in percentages of the
		fluid term: $100\times (\text{kin}-\text{fluid})/\text{fluid}$. A discrepancy
		of $100\%$ therefore corresponds
		a factor two between the quantities of interest. Note that the discrepancy is positive
		if the fluid term is smaller than the kinetic term.}
	\label{Fig:momentum_energy_discr_CX}
\end{figure}

We now turn our attention to charge exchange.
The fluid treatment of charge exchange has two main underlying
assumptions: the Maxwellianity of the plasma, and the linearity of
the momentum exchange.
The impact of the former heavily depends on the shape of $f_i$ and is
hard to generalize. Here we investigate the impact
of the latter on the momentum and energy exchange.
In that sense, we scan a range of fluid velocities for the ion species,
as well as a range of temperatures. We then compute the relative difference
between the fluid moment, Eq.~\eqref{eq:CX_sources_fluid}, and
the moments of the kinetic term, Eq.~\eqref{eq:gyro_term_cx} for
Maxwellian distribution functions.
We note that the fluid velocity of the plasma
is not taken into account when computing $\svcx$ \cite{reiter_2000},
leading to further discrepancies at high fluid velocity of the ions.
We do this comparison twice: one with the assumption
of cold neutrals ($T_N=4\eV$) and one with the assumption of
hot neutrals ($T_N=T_i$).
We plot the error relative to the fluid term, meaning $100\times
	(\text{kin}-\text{fluid})/\text{fluid}$.
The results are shown in Fig.~\ref{Fig:momentum_energy_discr_CX}.

Fig.~\ref{Fig:momentum_energy_discr_CX} shows only positive errors,
meaning that the fluid computation always underestimates the energy and momentum
lost by the ions when undergoing charge exchange reactions.
Moreover, we can see that the error is systematically higher for hot neutrals
compared to the one for cold neutrals.
This is due to the $\svcx$ used, which does not depend on the ion
fluid velocity, as would physically be required. When using $T_N=T_i$,
the difference in fluid velocity is the dominant effect, leading to large discrepancies.
However, even for hot neutrals, the discrepancy remains substantial,
up to $100\%$, meaning a factor $2$ between the fluid term and the kinetic one.
This discrepancy is maximal at low temperature and high fluid velocity, when
the dominant driver for momentum and energy transfer between ions and neutrals
is the difference of fluid velocities. With this, we conclude our comparison between
the kinetic model and fluid source terms.

\section{Conclusion}
\label{Sec:conclusion}

\subsection{Summary and conclusion}

In this work, we have derived comprehensive fully kinetic and gyrokinetic inelastic
collision operators that account for electron impact ionization, resonant charge
exchange and radiative recombination of hydrogen-like atoms acting on both the electron
and ion species. To model ionization, we have used simplified triply differential cross
sections. These keep the full energy dependence of the cross section but reduce the
angular distribution
to a single form for each of the two classes of outgoing electrons: a Dirac delta for
one and an isotropic distribution for the other. We have then incorporated radiation loss
and multi-step ionization through dressed cross sections and a radiation term.
The three kinetic operators then have been translated to a gyrokinetic framework,
accounting for the variation of impact energy along the ion and electron
gyromotion, but neglecting the spatial dependence created by their finite
Larmor radius (drift-kinetic approach).
The fluid moments of the newly derived model reduce to usual fluid source terms
for plasma-neutral interactions under the right set of fluid hypotheses.
Namely, all the distribution functions are taken to be Maxwellian, for ionization
and recombination the neutral velocity is assumed to be negligible and for charge
exchange the momentum transfer is assumed to be linear.

These gyrokinetic operators have been applied to both Maxwellian and non-Maxwellian
distribution functions for the plasma species. It has been found that, even when
taking Maxwellians for both the electron and the neutral species, the electron source
term exhibits marked features in velocity space, specific to kinetic and gyrokinetic
approaches, driving the distribution function away from a Maxwellian. For instance,
the ejection of cold electron by ionization
is sharply peaked around the origin, emitting electrons with a characteristic velocity
far lower than the electron thermal velocity. Moreover, it has been shown that, in most
cases, fluid hypotheses fail to predict the depletion and population of the distribution
function in velocity space. In particular, for velocities in the vicinity of the thermal
speed, fluid and kinetic approaches can predict opposite trends (a source in one is a sink
in the other). Such effects on the bulk of the plasma distribution functions in velocity space
may be particularly impactful on microturbulence instabilities, potentially altering their
drive and saturation. For charge exchange, we have studied how sensitive are
the momentum and energy transfer from ions to neutrals with respect to the assumption
of linearity of the operator. In that regard, we found discrepancies up to a factor 2
between gyrokinetic and fluid approaches, maximal at low temperatures and high fluid
velocities.

The kinetic and gyrokinetic formulations derived in this paper
provide detailed description of how non-elastic collisional processes
redistribute particles in velocity space. This processes, happening at time
scales comparable to both turbulent time scales and collision ones (depending
on the plasma and neutral conditions), can affect the dynamics of the plasma in unknown
ways. The interaction with plasma instabilities, resonant for certain regions of
the velocity space close to the thermal velocity, or with resonant heating, needs to
be studied. Moreover, the production of cold electrons in the SOL \textit{via} ionization,
and their trapping in open field lines between two sheath potential drops at the walls
could affect the equilibrium with collisions, which drive the distributions towards Maxwellians.

\subsection{Generalization to others atoms}

The derivation undergone in this article applies to hydrogen isotopes.
It is nevertheless possible to generalize to heavier atoms.
The challenge is threefold. Firstly, heavier atoms, like oxygen ($\mathrm{O}$),
have more than one ionization state, each of them with a different ionization energy.
Each one of these ionization levels leads to a different species to track in its
own right, with a different distribution function and charge.
Each of these species undergoes ionization
($\mathrm{O}$ to $\mathrm{O}^{+}$, $\mathrm{O}^{+}$ to $\mathrm{O}^{2+}$,...)
that needs to be solved kinetically. The time required to compute one of these
interactions can be large enough to prohibit the computation of many. This, however,
could be approximated by bundling of charge states. Charge exchange suffers from the
same problem, needing to be solved for each charge state. It is however
not resonant when occurring between different charge states, and does not need to be solved
for each pair of charge states.

Secondly, heavier atoms can undergo ionization from an inner shell, exhibit more excitation
states, and even metastable ones. This needs to be included in the ionization computation,
and would certainly prove
especially important when dressing the cross sections. The CRM is more useful than ever
when studying impurities, simplifying this question. However, it does not solve the problem
of the varying ionization energy, than needs to be carefully incorporated in the
radiation operator.

Thirdly, for all these atoms and their ionization states,
one needs to have all the experimental and/or analytical data, both for the differential
cross section in energy (not only taking into account outer shell ionization) and for
the angle distribution of outgoing electrons, depending on their energy. This latter
distribution is particularly important to check whether an assumption similar to
Eq.~\eqref{eq:ass_on_dsdedodo} could be made.

Finally, some words have to be said to speak about molecules. Throughout this article
the matter has been carefully avoided, and for good reasons. Kinetic studies of molecule
interactions is known to be a highly non-trivial. Incorporating molecules in
a plasma simulation greatly increases the number of processes to consider. An incomplete
list of them can be found in the so-called Kotov model, see \cite{kotov_2008}.
Moreover, molecules in a given vibrational state can undergo resonant charge exchange
with other ion species, and assist (as a transition state) atomic processes.
Quantitative predictions for experiments like ITER will likely not be able to overlook
this, the dynamics of neutrals under the ITER dome being believed to be dominated by
elastic collisions with molecules \cite{sureshkumar_2024}.
This aspect is well beyond the framework of the present manuscript whose goal was rather
to provide a sound basis for self-consistent gyrokinetic and fully kinetic calculations
of plasma interplay with hydrogen-like neutrals in the confined or near SOL region
of fusion devices.

\FloatBarrier

\begin{appendices}
	\section{Choices for the cross sections}
	\label{Annex:cross_sections}
	
\begin{figure}[ht]
	\centering
	\includegraphics[width=0.6\linewidth]{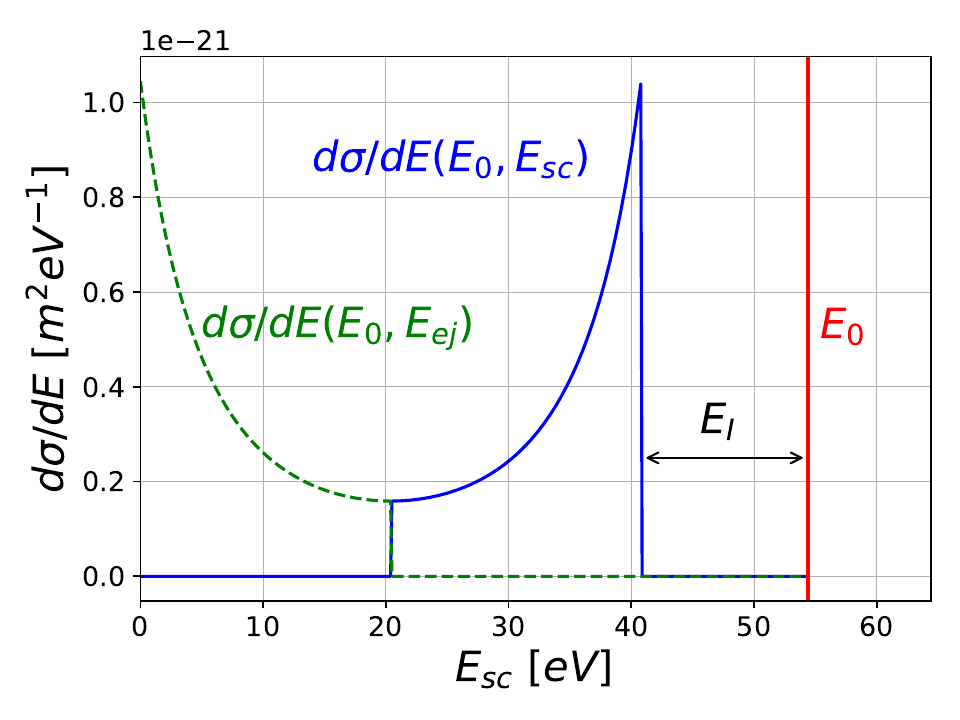}
	\caption{Splitting of the BEB differential cross section between
		ejected and scattered electrons}
	\label{fig:def_beb}
\end{figure}

	Historically, the ionization cross section (and differential
cross section) of a hydrogen
atom has been first derived in \cite{thomson_xlii._1912}
by Thomson in 1912. Although the derivation is
straightforward, because only incorporating classical physics,
the result is far from experimental observations.
Better approximations have been developed, such as the Mott
cross section \cite{czyzewski_1990}, or the BEB (binary-encounter-Bethe)
one \cite{kim_1994}, amongst others.
Most of these more refined approximations relies on quantum
computations. Therefore, they do not distinguish between the
scattered electron and the ejected one. This leads to symmetrical
expressions in energy, as shown by the blue and green curves in
Figure~\ref{fig:def_beb}
For instance, the BEB differential cross section, used in this paper, is
\begin{align}
	 & \left(\dsde\right)_\text{BEB}(\Ein, E)=
	\frac{4\sigma_B}{E_I}
	\left[ F_1(A_1 + A_2)
		+ F_2(A_1^2 + A_2^2)
		+ F_3(A_1^3 + A_2^3) \right]
	\label{eq:BEB_dsde}
	\\
	 & F_2 = \frac{1}{1 + u + t},
	\qquad F_1 = - \frac{F_2}{1 + t},
	\qquad F_3 = F_2\ln t,
	\qquad A_1 = \frac{1}{1 + w},
	\qquad A_2 = \frac{1}{t - w},
	\nonumber
\end{align}
with $t=\Ein/E_I$, $w=E/E_I$ and $u=U/E_I$ the normalised
kinetic energy of the bound electron, which by Virial theorem
we know $u=1$.
To use it in our framework, we need
to precise which one of the electrons is the scattered one.
This has been done in Section~\ref{Sec:iz_ass}, where we defined
the scattered electron to be the most energetic of the two outgoing
electrons. This is shown in Figure~\ref{fig:def_beb} by the
two curves in blue and dashed green, representing respectively
scattered and ejected electrons. Eq.~\eqref{eq:BEB_dsde} then becomes
\begin{align*}
	\left(\dsde\right)_\text{BEB}(\Ein, \Esc)
	 & =0
	\iff \Esc\notin\left[\frac{E_0-E_I}{2},E_0-E_I\right]
	\\
	\left(\dsde\right)_\text{BEB}(\Ein, \Eej)
	 & = \left(\dsde\right)_\text{BEB}(\Ein, E_0-\Eej-E_I)
\end{align*}

The ionization cross section $\si$ is then found by integrating
Eq.~\eqref{eq:BEB_dsde} in energy, yielding
\begin{equation*}
	\si^\text{BEB}(E_0) = 4\sigma_B \frac{1}{1 + u + t}
	\left[\frac{\log(t)[1-1/t^2]}{2t^2}
		+ 1 - \frac{1}{t} - \frac{\log t}{t + 1}
		\right].
\end{equation*}

\begin{figure}[ht]
	\centering
	\includegraphics[width=0.6\linewidth]{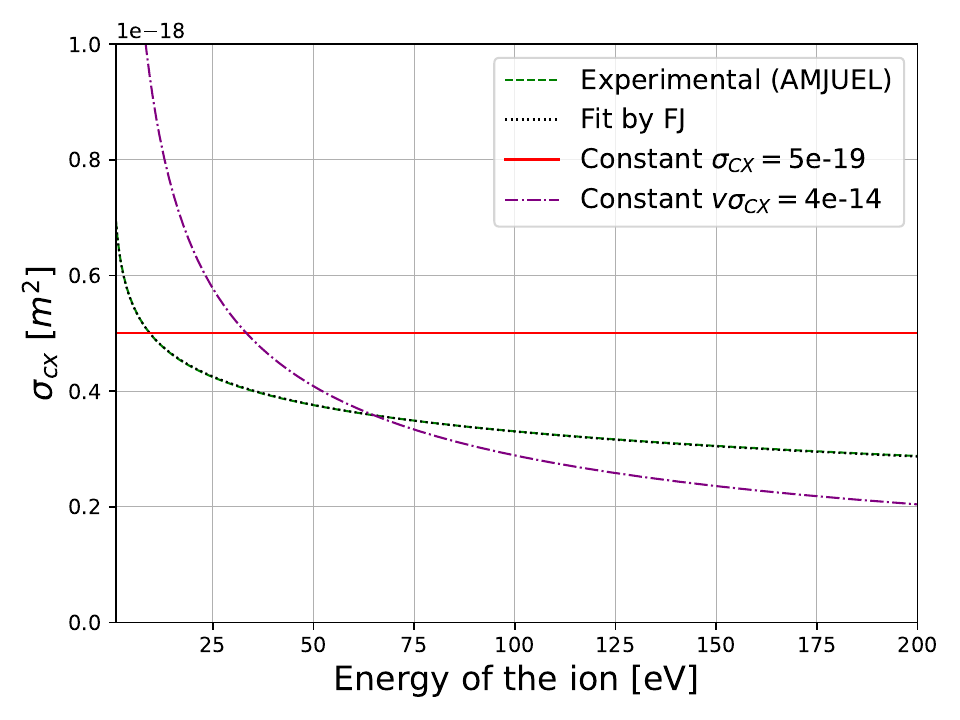}
	\caption{Data for $\scx$ (experimental and fitted) compared with a constant $\scx$
		and the value for $\scx$ giving and constant $v\scx(v)$.}
	\label{fig:cx_sigma}
\end{figure}

For charge exchange, the cross section can be found
in databases like AMJUEL.
A very good fit has been produced by Janev \cite{janev_1995}
and reads
\begin{equation}
	\scx(E) = 0.6937\times 10^{-18}
	\frac{[1 - 0.155 \log E]^2}{1 + 0.1112\times 10^{-14} E^{3.3}}.
\end{equation}
Fig.~\ref{fig:cx_sigma} shows a comparison of these two curves
with two widely used assumptions: $\scx$ constant and $v\scx(v)$ constant.
The assumption of constant $v\scx(v)$ arises from a linearity hypothesis for the
momentum exchange of charge exchange. If we want to have the momentum
of Eq.~\eqref{eq:termf_cx} to equate Eq.~\eqref{eq:CX_sources_fluid} for
Maxwellian distribution functions, we need to have
\begin{multline*}
	\int d^3\av\int d^3\aw\;
	m_i \av |\av-\aw| \scx\left(\frac{1}{2}m_i|\av-\aw|^2\right)
	\Max{i}(\av)\Max{N}(\aw)
	\\
	= \svcx \int d^3\av\; m_i \av\Max{i}(\av)
	\int d^3\aw\; \Max{N}(\aw),
\end{multline*}
where we only wrote one of the two terms of Eq.~\eqref{eq:termf_cx}.
This is true only if $|\av-\aw| \scx\left(\frac{1}{2}m_i|\av-\aw|^2\right)$
does not depend on $\av$ and $\aw$, and therefore is constant.

	\section{Full kinetic derivations}
	\label{Annex:kin_details}
	In this section we derive both Eq.~\eqref{eq:term_sd_source}
and \eqref{eq:termej} from Section~\ref{Sec:iz_ground_state}.
For the scattered electron, we start from Eq.~\eqref{eq:sd_no_density}
and inject Eq.~\eqref{eq:1_expansion_termE}.
We change twice variables, using $\au=\av_0-\aw$
(and therefore $\Ein=\frac{1}{2}m_e u^2$),
and $\Esc=\frac{1}{2}m_e u_{\sc}^2$, to obtain
\begin{multline}
	\int d^3 \avsc\;\termEe(\avsc)
	\\
	= \int d^3\aw \;f_N(\aw) \int d^3\au\;u \int d\Osc\;\delta_0(\Osc)
	\int_0^{u}du_\sc\; m_e u_\sc\dsde(\Ein,\Esc) f_e(\au+\aw).
	\label{eq:2_expansion_termE}
\end{multline}
We then continue by expressing $d^3\au$ in spherical coordinates,
$\au=u(\sin\theta\cos\varphi, \sin\theta\sin\varphi, \cos\theta)$.
Note that this change of variable introduces a $\sin\theta u^2$
Jacobian in the integral.
We furthermore take the time here to remark an important point.
The solid angle $\Osc$ has been defined as the angle between $\av_0$ and $\avsc$.
It is the same as the angle between $\au=\av_0-\aw$ and
$\au_\sc = u_\sc\Osc = \av_\sc-\aw$.
Moreover, we note that, in Eq.~\eqref{eq:2_expansion_termE},
inside of the $\Osc$ integral is a Dirac delta. This comes from
the assumption presented in Eq.~\eqref{eq:ass_on_dsdedodo}.
This Dirac delta enforces
the collinearity of $\avsc$ and $\av_0$, and therefore of
$\au_\sc$ and $\au$. This, in turn, yields
$\au_\sc=u_\sc(\sin\theta\cos\varphi, \sin\theta\sin\varphi, \cos\theta)$.
This is the key that allows us to exchange the two integral signs
in $u$ and $u_\sc$.
In the following we do not write the Dirac delta, but
it is to be remembered that it is hidden in the very definition
of $\au_\sc$. We write
\begin{multline}
	\int d^3 \avsc\;\termEe(\avsc)
	\\
	= \int d^3\aw \;f_N(\aw)
	\int d\varphi\int d\theta
	\int_0^{\infty}du_\sc
	\int_{u_\sc}^\infty du\;
	\sin\theta\ u^3
	m_e u_\sc \dsde(\Ein,\Esc)f_e(\au+\aw).
	\label{eq:3_expansion_termE}
\end{multline}
Now that we exchanged the two integral signs, we use our new
definition coming from Assumption~\eqref{eq:ass_on_dsdedodo}, namely
$\au_\sc=u_\sc(\sin\theta\cos\varphi, \sin\theta\sin\varphi, \cos\theta)$,
for a change of variable. Note that it introduces a Jacobian
in the integral, precisely $1/\sin\theta u_\sc^2$.
Moreover, $\au$ becomes $\au_\sc u/u_\sc$.
We then switch back to $\avsc=\au_\sc+\aw$,
\begin{multline}
	\int d^3 \avsc\;\termEe(\avsc)
	\\
	= \int d^3\avsc
	\int d^3\aw \;f_N(\aw)
	\int_{|\avsc-\aw|}^\infty du\;
	m_e\frac{u^3}{|\avsc-\aw|}
	\dsde(\Ein,\Esc)f_e\left(\aw + u \frac{\avsc-\aw}{|\avsc-\aw|}\right)
	\label{eq:4_expansion_termE}
\end{multline}
According to these computations, we
have a source term with the following form:
\begin{equation*}
	\termEe(\avsc)=
	\int d^3\aw \;f_N(\aw)
	\int_{|\avsc-\aw|}^\infty du\;
	\frac{m_eu^3}{|\avsc-\aw|}
	\dsde(\Ein,\Esc)f_e\left(\aw + u \frac{\avsc-\aw}{|\avsc-\aw|}\right).
\end{equation*}
Finally, to simplify the expression, we define a new dimensionless
quantity $\lsd$, as a ratio of velocities before
and after collision, presented in Eq.~\eqref{eq:def_lambda_sd}
(we recall here that $u=|\av_0-\aw|$).
This leads to Eq.~\eqref{eq:term_sd_source}.

For the ejection term, the derivation is similar,
starting from Eq.~\eqref{eq:ej_no_density}
and injecting Eq.~\eqref{eq:1_expansion_termej}.
We again use the change of variable $\au=\av_0-\aw$:
\begin{multline*}
	\int d^3 \avej\;\termej(\avej)
	\\
	= \int d^3\aw \;f_N(\aw) \int d^3\au\;u
	\int d\Oej\;\frac{1}{4\pi}
	\int_0^{\Ein}d\Eej\; \dsde(\Ein,\Ein-\Eej-E_I)f_e(\au+\aw).
\end{multline*}
We then use $\Eej=\frac{1}{2}m_eu_\ej^2$ for $u_\ej=|\au_\ej|=|\avej-\aw|$,
and the assumption that the ejected electron are uniformly
distributed in the solid angles $\Oej$. It allows us to move to a three-dimensional
integral in $\au_\ej$, inserting a $1/u_\ej^2$ Jacobian.
As before, we thus can exchange the two integral signs.
Lastly, we switch back to $\av_\ej=\au_\ej+\aw$ and
$\av_0=\au+\aw$:
\begin{align*}
	\int d^3
	 & \avej\;\termej(\avej)
	\\
	 & = \int d^3\aw \;f_N(\aw) \int d^3\au\;u
	\int_{u_\ej\le u}d^3\au_\ej\;\frac{m_e}{4\pi u_\ej}
	\dsde(\Ein,\Ein - \Eej - E_I)f_e(\au+\aw)
	\\
	 & = \int d^3\aw \;f_N(\aw)
	\int d^3\au_\ej
	\int_{u\ge u_\ej}d^3\au\;
	\frac{m_eu}{4\pi u_\ej}
	\dsde(\Ein,\Ein - \Eej - E_I)f_e(\au+\aw)
	\\
	 & = \int d^3\aw \;f_N(\aw)
	\int d^3\avej
	\int_{|\av_0-\aw|\ge|\avej-\aw|}d^3\av_0\;
	\frac{m_e|\av_0-\aw|}{4\pi |\av_\ej-\aw|}
	\dsde(\Ein,\Ein - \Eej - E_I)f_e(\av_0)
\end{align*}
We recover the term $\inlinetermej$, presented in Eq.~\eqref{eq:termej}.
This concludes the section.

	\section{Full gyrokinetic derivations}
	\label{Annex:gyro_details}
	In this section we start from Eq.~\eqref{eq:termf_cx}
and compute its gyrokinetic equivalent to obtain
Eq.~\eqref{eq:gyro_term_cx} and \eqref{eq:def_G_CX}.
The choice has been made to only present the computations
for charge exchange, the others being similar, to reduce redundancy.
We express $\av$ and $\aw$ in the parallel and perpendicular
directions:
\begin{equation*}
	\av=(v_\pll, v_\perp\cos\zeta, v_\perp\sin\zeta),
	\quad \aw=(w_\pll, w_\perp\cos\xi, w_\perp\sin\xi).
\end{equation*}
To shorten the notations we set
\begin{gather}
	\Emincx = \frac{1}{2}m_i \Big[|v_\pll-w_\pll|^2+|v_\perp-w_\perp|^2\Big],
	\label{eq:E_min_cx_annex}
\end{gather}
such that,
\begin{equation}
	\frac{1}{2}m_i|\av-\aw|^2
	=\Emincx+2m_i v_\perp w_\perp \sin^2\big(|\zeta-\xi|/2\big)
	= E^\cx\big(|\zeta-\xi|/2\big)
	\label{eq:impact_energy_cx_wrt_Emin_annex}
\end{equation}

We then express Eq.~\eqref{eq:termf_cx}
in cylindrical coordinates, using Eq.~\eqref{eq:E_min_cx_annex}
and \eqref{eq:impact_energy_cx_wrt_Emin_annex},
\begin{align*}
	   & \termf{i}{cx}(v_\pll, v_\perp, \zeta)
	\\
	{} & \qquad = {}
	\begin{multlined}[t][0.9\displaywidth]
		\int dw_\pll \int dw_\perp \;w_\perp \int_0^{2\pi} d\xi\;
		\scx\Big( E^\cx\big(|\zeta-\xi|/2\big) \Big)
		\sqrt{2E^\cx\big(|\zeta-\xi|/2\big)/m_i}
		\\
		\Big[f_N(v_\pll, v_\perp, \zeta) f_i(w_\pll, w_\perp, \xi)
			- f_i(v_\pll, v_\perp, \zeta) f_N(w_\pll, w_\perp, \xi)
			\Big].
	\end{multlined}
\end{align*}
To recover the effect of the ion gyromotion on this term, we need to average
over the gyrophase $\zeta$.
Moreover, we replace the particle distribution function $f_i$ by
the gyrocentre one $\of_i$.
\begin{align*}
	   & \termof{i}{cx}(v_\pll, v_\perp)
	\\
	{} & \qquad = \frac{1}{2\pi}\int_0^{2\pi}d\zeta\;
	\termf{i}{cx}(v_\pll, v_\perp, \zeta)
	\\
	{} & \qquad = {}
	\begin{multlined}[t][0.9\displaywidth]
		\frac{1}{2\pi}\int_0^{2\pi}d\zeta\;
		\int dw_\pll \int dw_\perp \;w_\perp \int_0^{2\pi} d\xi\;
		\scx\Big( E^\cx\big(|\zeta-\xi|/2\big) \Big)
		\sqrt{2E^\cx\big(|\zeta-\xi|/2\big)/m_i}
		\\
		\Big[f_N(v_\pll, v_\perp, \zeta) \of_i(w_\pll, w_\perp)
			- \of_i(v_\pll, v_\perp) f_N(w_\pll, w_\perp, \xi)
			\Big].
	\end{multlined}
\end{align*}
We can then use the change of variable $\varphi=\zeta-\xi$
(keeping respectively $\zeta$ and $\xi$ in each of the integrals),
along with the periodicity of $\sin^2$, and the fact that it is an even function.
\begin{align*}
	   & \termof{i}{cx}(v_\pll, v_\perp)
	\\
	{} & \qquad =
	\int dw_\pll \int dw_\perp \;w_\perp
	\of_i(w_\pll, w_\perp)
	\int_0^{2\pi} d\varphi\;
	\scx\Big( E^\cx\big(\varphi/2\big) \Big)
	\of_N(v_\pll, v_\perp)
	\sqrt{2E^\cx\big(\varphi/2\big)/m_i}
	\\
	{} & \qquad\qquad -
	\int dw_\pll \int dw_\perp \;w_\perp
	\of_i(v_\pll, v_\perp)
	\int_0^{2\pi} d\varphi\;
	\scx\Big( E^\cx\big(\varphi/2\big) \Big)
	\of_N(w_\pll, w_\perp)
	\sqrt{2E^\cx\big(\varphi/2\big)/m_i},
\end{align*}
where we used the definition presented in Eq.~\eqref{eq:def_average_fN}.
We define
\begin{align*}
	\Gcx(v_\pll,v_\perp, w_\pll, w_\perp)
	 & = \int_0^{2\pi} d\varphi\;
	\scx\Big( E^\cx\big(\varphi/2\big) \Big)
	\sqrt{2E^\cx\big(\varphi/2\big)/m_i}
	\\
	 & = 4\int_0^{\pi/2} d\theta\;
	\scx\Big( E^\cx(\theta) \Big)
	\sqrt{2E^\cx(\theta)/m_i}.
\end{align*}
It follows that $\Emincx$ is the only dependence in the parallel
velocities, and that it only depends on the difference between
$v_\pll$ and $w_\pll$. We can therefore omit one dependence
of $\Gcx$, using $\deltavpll=v_\pll-w_\pll$,
and find Eq.~\eqref{eq:gyro_term_cx}.

\end{appendices}


\bibliographystyle{plain}
\bibliography{biblio.bib}

\end{document}